\documentclass[12pt]{iopart}
\expandafter\let\csname equation*\endcsname\relax
\expandafter\let\csname endequation*\endcsname\relax

\usepackage[normalem]{ulem}
\usepackage{bm}
\usepackage[latin1]{inputenc}
\usepackage{textcomp}
\usepackage{mathptmx}
\usepackage{amsmath}
\usepackage[charter,cal=cmcal]{mathdesign}
\usepackage[usenames]{color}
\usepackage{graphicx}% Include figure files
\usepackage{dcolumn}% Align table columns on decimal point
\usepackage{bm}% bold math
\usepackage{hyperref}% add hypertext capabilities
\usepackage{color}
\usepackage{dsfont}
\usepackage{pgf}
\usepackage{pgfplots}
\usepackage{adjustbox}

\usepackage{physics}
\definecolor{rainbow5red}{RGB}{237, 63, 52}
\definecolor{rainbow5orange}{RGB}{244, 118, 57}
\definecolor{rainbow5yellow}{RGB}{253, 244, 153}
\definecolor{rainbow5green}{RGB}{123, 192, 67}
\definecolor{rainbow5blue}{RGB}{2, 146, 207}
\definecolor{lightgray}{RGB}{221,221,221}
\newcommand{\ham}{\hat{\mathcal{H}}}

\begin{document}

\title{Optimal quantum interferometry robust to detection noise using spin-1 atomic condensates}

\author{Artur Niezgoda}
\address{Faculty of Physics, University of Warsaw, ul. Pasteura 5, PL-02-093 Warsaw, Poland}
\author{Dariusz Kajtoch}
\address{Institute of Physics, PAS, Aleja Lotnik\'{o}w 32/46, PL-02-668 Warsaw, Poland}
\author{Joanna Dzieka\'nska}
\address{Institute of Physics, PAS, Aleja Lotnik\'{o}w 32/46, PL-02668 Warsaw, Poland}
\author{Emilia Witkowska}
\ead{ewitk@ifpan.edu.pl}
\address{Institute of Physics, PAS, Aleja Lotnik\'{o}w 32/46, PL-02-668 Warsaw, Poland}

\date{\today}

% ----------------------------------
% ABSTRACT
% ----------------------------------
\begin{abstract}
Implementation of the quantum interferometry concept to spin-1 atomic Bose-Einstein condensates is analyzed by employing a polar state evolved in time. In order to identify the best interferometric configurations, the quantum Fisher information is maximized. Three optimal configurations are identified, among which one was not reported in the literature yet, although it gives the highest value of the quantum Fisher information in experimentally achievable short time dynamics. Details of the most optimal configurations are investigated based on the error-propagation formula which includes the interaction-based readout protocol to reduce the destructive effect of detection noise. In order to obtain Heisenberg scaling accessible by present day experimental techniques, an efficient measurement and a method for the inversion of dynamics were developed, as necessary for the protocol's implementation.
\end{abstract}

\maketitle

\section{Introduction}
\label{sec:intro}
Quantum interferometry that initially emerged in the quantum optics domain a little while back was successfully applied to systems composed of massive particles.
Numerous proof-of-principle experiments have demonstrated potential of ultra-cold atoms in precision measurements based on interferometric techniques \cite{RevModPhys.90.035005}. 
Today, ultra-cold atoms play an important role in measurements of physical quantities that could not be measured with optical devices, or they are measured with weaker precision. Chip-scale inertial sensors for real-time positioning and navigation, ultra-precise atomic clocks or magnetometers operating in Earth's magnetic field are good examples \cite{doi:10.1063/1.2216932,Ockeloen2013,PhysRevLett.113.103004}.

Spinor Bose-Einstein condensates consist of atoms with the total spin $F$ occupying internal Zeeman states numerated by the quantum magnetic number $m_F = 0, \pm1 ,\cdots, \pm F$. The atoms are exposed to the same external trapping potential independent of their internal state~\cite{KAWAGUCHI2012253}. 
Among various sensors achievable with ultra-cold atoms, spinor Bose-Einstein condensates with Zeeman energy levels sensitive to magnetic field can be used to encode information about unknown physical quantities using quantum interferometry techniques \cite{PhysRevLett.117.143004,PhysRevLett.117.013001,PhysRevA.97.043813,PhysRevA.97.023616,PhysRevA.93.023627,PhysRevA.97.032339,Pezze2017}.
Furthermore, interactions between atoms allow generating non-classical states, such as squeezed or entangled states.
The two strategies for non-classical states generation were considered theoretically and experimentally with spin-1 condensates so far. In the first one, the states of interest are generated dynamically from an initial coherent state, e.g. the polar state, as realized experimentally~\cite{PhysRevLett.117.143004,PhysRevLett.117.013001,Hamley2012, PhysRevLett.112.155304, Lucke2011, PhysRevLett.111.180401}. In the second scheme, an adiabatic driving through quantum phase transitions generates the non-classical states what was theoretically considered~\cite{PhysRevA.97.032339,Pezze2017,PhysRevLett.111.180401} and experimentally demonstrated~\cite{Zou6381,Hoang9475,Luo620}.
The non-classical states used as an input state of a quantum interferometer allow  precision measurement below the shot-noise limit, potentially approaching the ultimate Heisenberg limit with highly entangled states. 
Utility of the non-classical states in quantum interferometry typically requires detection of particles with very low noise \cite{Demkowicz-Dobrzanski2012}, which is hardly achievable with atomic-based technology. 
Recently, the concept of interaction-based readout~\cite{PhysRevLett.116.053601,PhysRevLett.118.150401,PhysRevLett.119.193601,PhysRevA.94.010102,OberthalerSU11} were proposed, and verified experimentally~\cite{PhysRevLett.117.013001} for some special case, in order to overcome the detection noise problem.
The interaction-based readout is nothing else but a unitary evolution applied to the quantum interferometer after the phase encoding step, but before the measurement takes place. Typically, the unitary evolution is based on the inter-particle interactions, the same as used for the non-classical state preparation. 

The purpose of this paper is to perform comprehensive study of quantum interferometry using spin-1 atomic Bose-Einstein condensates. We consider the non-classical states of the system generated from the initial polar state in the absence of an external magnetic field. The specific configuration we focus on was studied in this context, and the possibility of metrological gain was demonstrated experimentally for two different interferometric configurations~\cite{PhysRevLett.117.143004, PhysRevLett.117.013001, Lucke2011, OberthalerSU11}. On the theoretical level, however, it is interesting to prove and explain which configuration is the most optimal one, i.e. gives the highest possible and practicable precision. We believe that such analysis would help in understanding very foundations of quantum interferometry using spin-1 atomic condensates and further planning of experiments. In what follows, we consider the most general form of linear quantum interferometry \cite{PhysRevA.97.023616}. We identify optimal configurations of interferometric rotations by studying the quantum Fisher information (QFI) for the initial polar state. Our results show that the choices of mentioned experiments \cite{OberthalerSU11, Lucke2011} lie among the optimal once, however we found another configuration which determines the highest value of the QFI in short time dynamics accessible by nowadays experiments. We discuss how to achieve experimentally the best configuration taking into account the interaction-based readout to protect against detection noise. Finally, we show how that protocol, which is reversed evolution applied after the phase encoding step, can be realized experimentally with a single rotation of the state in the early evolution of the system.

The paper is organized as follows. In Section~\ref{sec:system} we present the model and show the analytical solution for the polar state $|0,N,0\rangle$ evolution, as well as other quantities important in derivation of the QFI value. In Section \ref{sec:identification} we calculate the QFI values and identify the best interferometric configuration. Next, in Section \ref{sec:observable} we discuss the effect of detection noise for various signals and show how the interaction-based readout can be implemented in the most optimal configuration in order to achieve the highest precision.

%%%%%%%%%%%%%%%%%%%%%%%%%%%%%%%%%%%%%%%%%%%%%%%%%%%%%%%%%%%%%%%%%%%%%%%%%%%%%%%%%%%%%%%%%%%%%%%%%
\section{The model and time evolution}
\label{sec:system}

We consider a spinor Bose-Einstein condensate, with three internal levels, in the single mode approximation (SMA) where all atoms from different Zeeman states occupy the same spatial mode $\phi(\mathbf{r})$, which satisfies the Gross-Pitaevskii equation with chemical potential $\mu$, see~\ref{app:SMA} for an explanation. We assume that the many-body Hamiltonian has the following form~\cite{KAWAGUCHI2012253, RevModPhys.85.1191, phdHamley}:
\begin{equation}\label{eq:ham}
\ham = \mu \hat{N} - c'_{0}\hat{N}(\hat{N}-1) + c'_2 (\hat{J}^2 - 2 \hat{N} ) + p\hat{J}_z + q \hat{N}_{s},
\end{equation} 
where $2c'_i = c_i \int d^3r |\phi(\mathbf{r})|^4$, $\hat{N}_{s} = \hat{N}_{+1} + \hat{N}_{-1}$, $\hat{N}_{m_F}$ is the operator of atoms number in the Zeeman state $m_F$ and $\hat{J}$ is the collective pseudo-spin operator defined within the SU(3) Lie algebra generators in the next section. 
The $c_0$ and $c_2$ coefficients can be expressed in terms of
s-wave scattering lengths~\cite{Ho1998,Machida}.
The last two terms in (\ref{eq:ham}) are linear and quadratic Zeeman energy shifts, respectively. 
The coefficient of the linear Zeeman energy shift is $p=g_J\mu_B B$, where $\mu_B$ is the Bohr magneton and $g_J$ is the gyromagnetic ratio. The coefficient of the quadratic Zeeman term may have two contributions from the external magnetic field ($q_B$) and from the microwave or light field ($q_{MW}$), therefore $q=q_B + q_{WM}$ [5]. The part controlled by the magnetic field is $q_B=(\mu_B B)^2/(4 E_{\rm HFS})$, where $E_{\rm HFS}$ is the hyperfine energy splitting. The value and the sign of $q_{MW}$ can be tuned independently of $q_B$ by employing a microwave field that is off-resonant with the other hyperfine state \cite{PhysRevA.73.041602}.
The Hamiltonian can be engineered in $F=1$~\cite{Hamley2012,Gerving2012,PhysRevLett.117.143004,Lange416,PhysRevLett.112.155304} or $F=2$~\cite{Lucke2011,PhysRevLett.117.013001} hyperfine manifold using ${}^{87}$Rb or ${}^{23}$Na atoms. The characteristic feature of the Hamiltonian~\eqref{eq:ham} is the conservation of the z-component of the collective pseudo-spin operator $[\ham,\hat{J}_z] = 0$. The Hamiltonian has a block-diagonal structure with each block labeled by the magnetization $M = -N, -N+1, \ldots, N$, which is the eigenvalue of the $\hat{J}_z$ operator. The linear Zeeman energy shift becomes irrelevant as it is proportional to the conserved magnetization, and the quadratic Zeeman energy is only important.  The value and sign of the quadratic Zeeman energy, through $q$, can be controlled using the magnetic field $B$ or the microwave dressing~\cite{PhysRevA.73.041602, PhysRevLett.115.163002,PhysRevA.65.033619}.

We start the evolution from the polar state $|0,N,0\rangle$, which is a ground state of the Hamiltonian~\eqref{eq:ham} in the high magnetic field limit. Since the initial state has $M=0$ and the quadratic Zeeman effect can be compensated with microwave dressing, the time evolution is governed by the Hamiltonian
\begin{equation}
\ham = c'_2 \hat{J}^2,
\end{equation} 
were we skipped constant terms, and is given by
\begin{equation}\label{eq:state_evolution}
|\psi(\bar{t})\rangle = e^{-i \bar{t} \hat{J}^2}|0,N,0\rangle,
\end{equation}
with $\bar{t} = t c'_2/\hbar$ which we assume to be positive,  as for e.g. ${}^{23}$Na atoms. The corresponding time scale is discussed in ~\ref{app:SMA}.

Since the $\hat{J}^2$ operator is diagonal in the total spin momentum basis, i.e. $\hat{J}^2 |N,J,M\rangle = J(J+1)|N,J,M\rangle$, it is convenient to decompose the polar state $|0,N,0\rangle$ into the total spin eigenbasis and then solve the evolution~\eqref{eq:state_evolution} analytically, which gives
\begin{equation}\label{eq:evolution_spin_basis}
	|\psi(\bar{t})\rangle = \sum_{J=0}^N {}^{'} D_{0}(N,J,0) e^{-i\bar{t}J(J+1)}|N,J,0\rangle,
\end{equation}
where 
\begin{equation}
	D_{0}(N,J,0) = \sqrt{\binom{\frac{N+J}{2}}{J} \binom{N+J}{J}^{-1} \frac{(2J+1)}{(N+J+1)}2^J},\label{eq:CJ}
\end{equation}
for zero magnetization, and the prim after the sum notation $\sum{}^{'}$ indicates summation over even values of $J=0, 2, 4, \dots, N$ when the total number of atoms $N$ is even, or summation over odd values of $J=1, 3, 5, \dots, N$ when $N$ is odd, see \ref{app:evolutionSPIN} for explanation. The representation~\eqref{eq:evolution_spin_basis} clearly demonstrates that the evolution is periodic with $\Delta\bar{t} = \pi$, more precisely $\ket{\psi(\bar{t})}=\ket{\psi(\bar{t}+n\pi)}$ where $n$ is an integer.

As the magnetization is conserved by the Hamiltonian, the time evolution of all quantities of interest considered in the next section can be expressed in terms of the following terms $\langle\hat{N}_{0}\rangle$, $\langle\hat{N}_{0}^2\rangle$ and $\langle \hat{a}^\dagger{}^2_0 \hat{a}_{1}\hat{a}_{-1} \rangle$. The evolution of those quantities can be calculated for the state (\ref{eq:evolution_spin_basis}) and they are~\footnote{In fact, to obtain the forms (\ref{eq:N0})-(\ref{eq:Imaa0apm}) it is more convenient to use the general representation of the spin state (\ref{eq:spin_state}) for $M=0$, use commutation relations listed in \ref{app:useful} to rearrange the order of operators and after some algebra get the results.}
\begin{eqnarray}
\langle\hat{N}_{0}\rangle  &=& N-4\sum_{J=0}^{N}{}^{'}{H_{1}(J)}\sin^2\left[\bar{t}(2J-1) \right], \label{eq:N0} \\
\langle\hat{N}_{0}^{2}\rangle & = & N^{2} - \sum_{J=0}^{N}{}^{'} H_{2}(J) \sin^2 \left[ 2\bar{t}(2J+5)\right] -  \sum_{J=0}^{N}{}^{'} H_{3}(J)\sin^2\left[ \bar{t}(2J-1)\right],
\end{eqnarray}
where
\begin{align}
&H_{1}(J) =\frac{J(J-1)N!}{(2J-1)(N-J)!!(N+J-1)!!}, \\
&H_{2}(J)=\frac{2^{J+4}N!(\frac{N+J}{2})!\prod_{v=1}^{4}(J+v)}{(\frac{N-J-4}{2})!(N+J+1)!\prod_{v=1}^3(2(J+v)+1)} ,\\
&H_{3}(J)=\frac{8(J^{2}-J-3+N(2J^{2}-2J-7))}{(2J-5)(2J+3)}H_1(J),
\end{align}
and
\begin{align}
&{\rm Re}   \langle \hat{a}^\dagger{}^2_0 \hat{a}_{1}\hat{a}_{-1} \rangle  
=\frac{1}{2} \left( N-\langle\hat{N}_0\rangle - 2 N \langle\hat{N}_0\rangle + 2 \langle\hat{N}_0^2\rangle  \right), \\
&{\rm Im}   \langle \hat{a}^\dagger{}^2_0 \hat{a}_{1}\hat{a}_{-1} \rangle    =\sum_{J=0}^{N-2}{}^{'}\frac{(J+1)(J+2) N!}{(N-J-2)!!(N+J+1)!!} \sin\left[ 2\bar{t}(2J+3)\right].\label{eq:Imaa0apm}
\end{align}
In addition, it is convenient to use also $\langle \hat{J}^2 \rangle = 2 N $, $\langle \hat{J}_z^2 \rangle = 0$. 

The evolution of the state (\ref{eq:evolution_spin_basis}), as well as above quantities, are usually considered using the recursion relation~\cite{PhysRevLett.81.5257}. Here, based on the decomposition of the polar state into the total spin eigenbasis similarly as in \cite{PhysRevA.54.4534}, we obtained quite simple to calculate analytical expressions.

%%%%%%%%%%%%%%%%%%%%%%%%%%%%%%%%%%%%%%%%%%%%%%%%%%%%%%%%%%%%%%%%%%%%%%%%%%%%%%%%%%%%%%%%%%%%%%%%%
\section{Identification of the best interferometeric configurations}\label{sec:identification}

%-----------------
\begin{figure}[]
	\centering
	\includegraphics[width=0.8\linewidth]{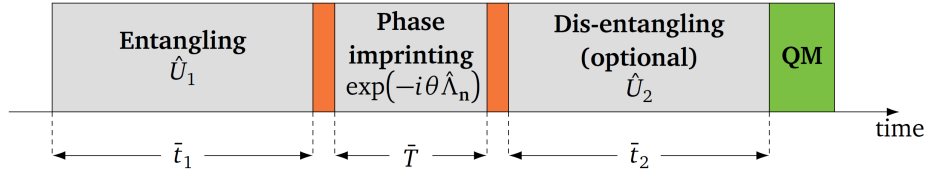}
\caption{General protocol for linear entanglement-enhanced quantum interferometry.}
\label{fig:fig1}
\end{figure}
%------------------

The interferometric protocol we consider consists of four steps in general, see Fig.\ref{fig:fig1}. The scheme starts with the dynamical state preparation by the unitary evolution $\hat{U}_1=e^{-i \bar{t}_1 \hat{J}^2}$ followed by the phase $\theta$ accumulation $\exp(-i\theta \hat{\Lambda}_{\mathbf{n}})$ during an interrogation time $\bar{T}$ under generalized generator of interferometric rotation $\hat{\Lambda}_{\mathbf{n}}$. The phase $\theta$ depends on the physical parameter to measure, e.g. magnetic field, and we assume that it is imprinted onto the state in the most general way. Next, an optional unitary evolution through the operator $\hat{U}_2$ can be applied before performing a quantum measurement (QM). In this section $\hat{U}_2=\mathds{1}$, however it is non-zero and plays a significant role in the interaction-based readout protocol considered in the next section. 

The purpose of this section is to identify the operator $\hat{\Lambda}_{\mathbf{n}}$ which determines the best precision in the $\theta$ estimation for the state (\ref{eq:state_evolution}) at the time $\bar{t}_1$.

In a general linear interferometer, the output state $|\psi(\theta)\rangle$ can be written as the action of the SU(3) rotation on the input state $|\psi(\bar{t}_1)\rangle$, i.e.
\begin{equation}
|\psi(\theta)\rangle = e^{-i\theta \hat{\Lambda}_{\mathbf{n}}}|\psi(\bar{t}_1)\rangle,
\end{equation}
where $\hat{\Lambda}_{\mathbf{n}}=\hat{{ \Lambda}} \cdot {\bf n}$ is a scalar product of a unit vector ${\bf n}$ and the vector $\hat{{ \Lambda}}=\{\hat{J}_x, \hat{Q}_{yz}, \hat{J}_y, \hat{Q}_{zx}, \hat{D}_{xy}, \hat{Q}_{xy},\hat{Y},\hat{J}_z\}$ composed of bosonic SU(3) Lie algebra generators:
 \begin{align}
 \hat{J}_{x} &\ =\ \frac{1}{\sqrt{2}}\left( \hat{a}^{\dag}_{\scriptscriptstyle{-1}}\hat{a}_{\scriptscriptstyle{0}} + \hat{a}^{\dag}_{\scriptscriptstyle{0}}\hat{a}_{\scriptscriptstyle{-1}} + \hat{a}^{\dag}_{\scriptscriptstyle{0}}\hat{a}_{\scriptscriptstyle{+1}} + \hat{a}^{\dag}_{\scriptscriptstyle{+1}}\hat{a}_{\scriptscriptstyle{0}}\right), \\
% %
 \hat{Q}_{zx} &\ =\ \frac{1}{\sqrt{2}}\left( -\hat{a}^{\dag}_{\scriptscriptstyle{-1}}\hat{a}_{\scriptscriptstyle{0}} - \hat{a}^{\dag}_{\scriptscriptstyle{0}}\hat{a}_{\scriptscriptstyle{-1}} + \hat{a}^{\dag}_{\scriptscriptstyle{0}}\hat{a}_{\scriptscriptstyle{+1}} + \hat{a}^{\dag}_{\scriptscriptstyle{+1}}\hat{a}_{\scriptscriptstyle{0}}\right), \\
% %
 \hat{J}_{y} &\ =\ \frac{i}{\sqrt{2}}\left( \hat{a}^{\dag}_{\scriptscriptstyle{-1}}\hat{a}_{\scriptscriptstyle{0}} - \hat{a}^{\dag}_{\scriptscriptstyle{0}}\hat{a}_{\scriptscriptstyle{-1}} + \hat{a}^{\dag}_{\scriptscriptstyle{0}}\hat{a}_{\scriptscriptstyle{+1}} - \hat{a}^{\dag}_{\scriptscriptstyle{+1}}\hat{a}_{\scriptscriptstyle{0}}\right), \\
% %
 \hat{Q}_{yz} &\ =\ \frac{i}{\sqrt{2}}\left( -\hat{a}^{\dag}_{\scriptscriptstyle{-1}}\hat{a}_{\scriptscriptstyle{0}} + \hat{a}^{\dag}_{\scriptscriptstyle{0}}\hat{a}_{\scriptscriptstyle{-1}} + \hat{a}^{\dag}_{\scriptscriptstyle{0}}\hat{a}_{\scriptscriptstyle{+1}} - \hat{a}^{\dag}_{\scriptscriptstyle{+1}}\hat{a}_{\scriptscriptstyle{0}}\right), \\
% %
 \hat{D}_{xy} &\ =\ \hat{a}^{\dag}_{\scriptscriptstyle{-1}}\hat{a}_{\scriptscriptstyle{+1}} + \hat{a}^{\dag}_{\scriptscriptstyle{+1}}\hat{a}_{\scriptscriptstyle{-1}}, \\
% %
 \hat{Q}_{xy} &\ =\ i\left( \hat{a}^{\dag}_{\scriptscriptstyle{-1}}\hat{a}_{\scriptscriptstyle{+1}} - \hat{a}^{\dag}_{\scriptscriptstyle{+1}}\hat{a}_{\scriptscriptstyle{-1}}\right), \\
% %
 \hat{Y} &\ =\ \frac{1}{\sqrt{3}}\left( \hat{a}^{\dag}_{\scriptscriptstyle{-1}}\hat{a}_{\scriptscriptstyle{-1}} - 2\hat{a}^{\dag}_{\scriptscriptstyle{0}}\hat{a}_{\scriptscriptstyle{0}} + \hat{a}^{\dag}_{\scriptscriptstyle{+1}}\hat{a}_{\scriptscriptstyle{+1}}\right),\\
% %
 \hat{J}_{z} &\ =\  \hat{a}^{\dag}_{\scriptscriptstyle{+1}}\hat{a}_{\scriptscriptstyle{+1}} - \hat{a}^{\dag}_{\scriptscriptstyle{-1}}\hat{a}_{\scriptscriptstyle{-1}} ,
 \end{align}
 where $\hat{a}_{m_F}$ is the annihilation operator of the particle in the $m_F$ Zeeman component. 

In this scheme, the minimal possible uncertainty of the parameter $\theta$ is determined by the inverse of the quantum Fisher information $\Delta \theta \geqslant 1/\sqrt{F_Q[|\psi(\bar{t}_1) \rangle, \hat{\Lambda}_{\mathbf{n}}]}$, which depends on the input state $|\psi(\bar{t}_1)\rangle$ and the generator of the interferometric rotation $\hat{\Lambda}_{\mathbf{n}}$. We will refer the generator of the interferometric rotation $\hat{\Lambda}_{\mathbf{n}}$ as an interferometer, for simplicity. 
The QFI is defined as~\cite{Smerzi}
\begin{equation}\label{eq:fisher_information}
F_{Q}[\hat{\rho}, \hat{\Lambda}_{\mathbf{n}}] = 4\mathbf{n}^{T} \cdot \Gamma[\hat{\rho}] \cdot \mathbf{n},
\end{equation}
where $\Gamma[|\psi (\bar{t}_1) \rangle]$ is the covariance matrix
\begin{align}\label{eq:covariance_matrix}
 \Gamma_{ij} = \frac{1}{2}\langle\hat{\Lambda}_i \hat{\Lambda}_j + \hat{\Lambda}_j \hat{\Lambda}_i \rangle - \langle \hat{\Lambda}_i \rangle \langle \hat{\Lambda}_j \rangle.
\end{align}
The maximal value of the QFI is given by the largest eigenvalue $\lambda_{\rm max}$ of the covariance matrix, and for the three level system considered here it is $F_Q = 4\lambda_{\rm max}$. The maximal possible value of the QFI is $F_Q = 4N^2$ and sets the Heisenberg limit for the estimation precision $\Delta \theta$, which can be attained only by the fully particle entangled states. On the other hand, separable states can give at most $F_Q=4N$ \cite{PhysRevA.98.013610}. The generator of the optimal interferometric rotation is determined by the eigenvector corresponding to the maximal eigenvalue of the covariance matrix (\ref{eq:covariance_matrix}). In general, the symmetric matrix (\ref{eq:covariance_matrix}) has 36 distinct elements, but for the input state $|\psi(\bar{t}_1)\rangle$ defined in the equation~\eqref{eq:state_evolution} most of its entries are $0$ due to rotational symmetry $e^{-i\alpha \hat{J}_z}|\psi(\bar{t}_1)\rangle = |\psi(\bar{t}_1)\rangle$, which holds for any $\alpha \in \mathds{R}$ due to conservation of magnetization. This property results in the block diagonal structure of the covariance matrix in the subspace of zero magnetization, which is the following:
\begin{equation}\label{eq: covariance}
\Gamma = \Gamma_{+} \oplus \Gamma_{-} \oplus [\Gamma_{55}] \oplus [\Gamma_{55}] \oplus [\Gamma_{77}] \oplus [0],
\end{equation}
where 
\begin{equation}
\Gamma_{+} = 
\left( 
	\begin{array}{cc}
    	\Gamma_{11} & \Gamma_{12} \\
        \Gamma_{12} & \Gamma_{22}
	\end{array}
\right),
\ \ \ 
\Gamma_{-} = 
\left( 
	\begin{array}{cc}
    	\Gamma_{11} & -\Gamma_{12} \\
        -\Gamma_{12} & \Gamma_{22}
	\end{array}
\right),
\end{equation}
and 
%based on time evolution (\ref{eq:evolution_spin_basis}) and results presented in the previous section one can write explicitly analytic expressions for
\begin{align}
	\Gamma_{11} & =  \langle \hat{J}_x^2 \rangle - \langle \hat{J}_x \rangle^2  = N, \\
	\Gamma_{22} &= \langle \hat{Q}_{yz}^2 \rangle - \langle \hat{Q}_{yz} \rangle^2= \langle \hat{N}_0(2\hat{N}_s + 1)\rangle , \\
	\Gamma_{12} & = \frac{1}{2}\langle \{\hat{J}_x, \hat{Q}_{yz} \} \rangle - \langle \hat{J}_x \rangle \langle \hat{Q}_{yz} \rangle =4 {\rm Im}   \langle \hat{a}^\dagger{}^2_0 \hat{a}_{1}\hat{a}_{-1} \rangle, \\
	\Gamma_{55} & =  \langle \hat{D}_{xy}^2 \rangle - \langle \hat{D}_{xy} \rangle^2 = \frac{1}{2}\langle \hat{N}_s (\hat{N}_s + 2) \rangle, \\
	\Gamma_{77} &= \langle \hat{Y}^2 \rangle - \langle \hat{Y} \rangle^2 = 3 \left( \langle \hat{N}_0^2 \rangle - \langle \hat{N}_0 \rangle^2 \right),
\end{align}
with averages of particular operators taken at $\bar{t}=\bar{t}_1$,~i.e. $\langle \cdot \rangle = \langle \psi(\bar{t}_1) | \cdot|\psi(\bar{t}_1) \rangle $.
Matrices $\Gamma_{\pm}$ share the same eigenvalues, but they have different eigenvectors~\footnote{If $(\hat{J}_x + \gamma \hat{Q}_{yz})/\sqrt{1+\gamma^2}$, with $2\Gamma_{12}\gamma=\Gamma_{22}-\Gamma_{11}+\sqrt{4\Gamma_{12}^2+(\Gamma_{11}-\Gamma_{22})^2}$, is an eigenvector of the matrix $\Gamma_{+}$, the $(\hat{J}_y - \gamma \hat{Q}_{zx})/\sqrt{1+\gamma^2}$ is an eigenvector of the matrix $\Gamma_{-}$, corresponding to the same eigenvalue. }. Due to the rotational symmetry we have $\Gamma_{33}=\Gamma_{11}$, $\Gamma_{44}=\Gamma_{22}$, $\Gamma_{66}=\Gamma_{55}$ and $ \Gamma_{34} =- \Gamma_{12}$~\cite{PhysRevA.97.023616}.

\begin{figure}[]
\centering
\includegraphics[width=0.6\linewidth]{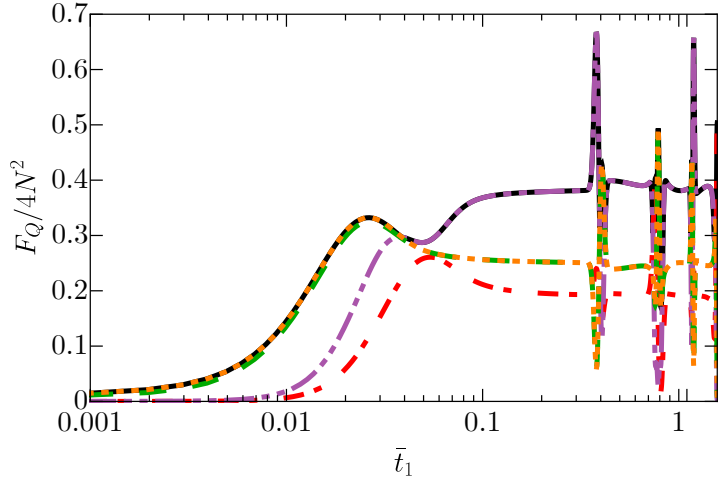}
\caption{Variations of the QFI in time for different interferometers $\hat{\Lambda}^{(I)}_{\mathbf{n}}= \hat{Q}_{yz}$ (dashed green line), $\hat{\Lambda}^{(II)}_{\mathbf{n}}=\hat{D}_{xy}$ (dot dashed red line) and $\hat{\Lambda}^{(III)}_{\mathbf{n}}=\hat{Y}$ (double dot dashed violet line). In addition, $\hat{\Lambda}^{(I)}_{\mathbf{n}}=\hat{J}_x(\gamma)$, with $\gamma$ chosen such that it maximizes the QFI value, is also shown by the orange dotted line for comparison. The  QFI maximized over all interferometers is shown by the black solid line. The evolution extends from $\bar{t}_1 = 0$ up to $\bar{t}_1=\pi/2$ as the rest can be recreated through reflections. The QFIs have a characteristic plateau region, where their values are stable for a long period of time and reveal Heisenberg scaling.
Further dynamics provides much more metrologically useful states, however the longer times regime is not accessible by present day experiments. The number of particles is $N=100$. Note the logarithmic scale on the horizontal axis.}
\label{fig:fig2}
\end{figure}

In general, the values of the QFI are determined by eigenvalues of the covariance matrix while the corresponding interferometric generators are set by the scalar product of the appropriate eigenvector of the covariance matrix and the vector $\hat{\Lambda}$. There are three different non-zero eigenvalues of the covariance matrix (\ref{eq: covariance}), and hence, three different eigenvectors which define three generators of interferometric rotation of practical importance. The generator of the interferometric rotation corresponding to the first non-zero  eigenvalue of the covariance matrix
%first eigenvector 
is
\begin{equation}
\hat{\Lambda}^{(I)}_{\mathbf{n}}  = \sqrt{\epsilon}\hat{J}_x(\gamma) + \sqrt{1-\epsilon}\hat{J}_y(\gamma) , \ \ \epsilon \in [0,1],
\end{equation}
where $\hat{J}_x(\gamma) = (\hat{J}_x + \gamma \hat{Q}_{yz})/\sqrt{1+\gamma^2}$ and $\hat{J}_y(\gamma) = (\hat{J}_y - \gamma \hat{Q}_{zx})/\sqrt{1+\gamma^2}$.
The value of $\epsilon$ does not change the value of the resulting QFI, and we will always consider $\epsilon=1$. The application of $\hat{\Lambda}^{(I)}_{\mathbf{n}}$ as a generator of interferometric rotation gives $F_Q^{(I)}=4 \Delta^2 \hat{J}_x(\gamma) $. We have checked numerically that when $\gamma \gg 1$, then $\hat{\Lambda}^{(I)}_{\mathbf{n}}\approx \hat{Q}_{yz}$ and $F_Q^{(I)}\approx 4\Gamma_{22}=4 \Delta^{2} \hat{Q}_{yz} $ as can be seen in Fig.\ref{fig:fig2}. Therefore, in the further part of the paper we will always take $\hat{\Lambda}^{(I)}_{\mathbf{n}} = \hat{Q}_{yz}$.
The generator of the interferometric rotation corresponding to the second non-zero eigenvalue of (\ref{eq: covariance}) is 
\begin{equation}
\hat{\Lambda}^{(II)}_{\mathbf{n}}  = \sqrt{\epsilon}\hat{D}_{xy} + \sqrt{1-\epsilon} \hat{Q}_{xy} , \ \ \epsilon \in [0,1], 
\end{equation}
and its usages as an interferometer will results in $F_Q^{(II)}=4\Gamma_{55}=4 \Delta^2 \hat{D}_{xy}$, for $\epsilon=1$. The generator of the interferometric rotation corresponding to the third non-zero eigenvalue of the covariance matrix is
\begin{align}
\hat{\Lambda}^{(III)}_{\mathbf{n}} & = \hat{Y} ,
\end{align}
and this interferometer turns out to give the QFI equal to $4\Gamma_{77}$, i.e. $F_Q^{(III)}=4 \Delta^2 \hat{Y}$. The optimal interferometers recognited by us have a two-mode nature~\cite{PhysRevA.98.013610}, as:
\begin{align}
	\hat{\Lambda}^{(I)}_{\mathbf{n}} & = i( \hat{g}_S\hat{a}^{\dagger}_0 - \hat{g}^{\dagger}_S\hat{a}_0), \\
	\hat{\Lambda}^{(II)}_{\mathbf{n}} & =  \hat{a}^{\dagger}_{-1}\hat{a}_{+1} + \hat{a}^{\dagger}_{+1}\hat{a}_{-1}, \\
	\hat{\Lambda}^{(III)}_{\mathbf{n}} & = \sqrt{3}(\hat{N}_{+1} + \hat{N}_{-1}) - 2\hat{N}/\sqrt{3},
\end{align}
where $\hat{g}_S = (\hat{a}_{+1} + \hat{a}_{-1})/\sqrt{2}$
~\footnote{The SU(3) algebra generators (14)-(21) can be written in terms of the symmetric $\hat{g}_S = (\hat{a}_{+1} + \hat{a}_{-1})/\sqrt{2}$ and anti-symmetric $\hat{g}_A = (\hat{a}_{+1} - \hat{a}_{-1})/\sqrt{2}$ bosonic operators and they are:
$\hat{J}_x=\hat{a}^{\dagger}_0 \hat{g}_S + \hat{a}_0 \hat{g}^{\dagger}_S$, $\hat{Q}_{zx}=\hat{a}^{\dagger}_0 \hat{g}_A + \hat{a}_0 \hat{g}^{\dagger}_A$, $\hat{J}_y=i(\hat{a}^{\dagger}_0 \hat{g}_A - \hat{a}_0 \hat{g}^{\dagger}_A)$, $\hat{Q}_{yz}=i(\hat{a}^{\dagger}_0 \hat{g}_S - \hat{a}_0 \hat{g}^{\dagger}_S)$,
$\hat{D}_{xy}=\hat{g}^\dagger_S \hat{g}_S - \hat{g}^\dagger_A \hat{g}_A$, 
$\hat{Q}_{xy}=i(\hat{g}^\dagger_S \hat{g}_A - \hat{g}^\dagger_A \hat{g}_S)$, $\hat{Y}=\frac{1}{\sqrt{3}}(\hat{g}^\dagger_S \hat{g}_S + \hat{g}^\dagger_A \hat{g}_A - 2 \hat{a}_0^\dagger \hat{a}_0))$,
$\hat{J}_{z}=\hat{g}^\dagger_S \hat{g}_A + \hat{g}^\dagger_A \hat{g}_S$.}
. In the case of $\hat{\Lambda}^{(I)}_{\mathbf{n}}$ the two modes are $\hat{a}_0$ and the symmetric $\hat{g}_S$ one. On the other hand, in the case of $\hat{\Lambda}^{(I)}_{\mathbf{n}}$ or $\hat{\Lambda}^{(III)}_{\mathbf{n}}$ they are always $\hat{a}_{\pm 1}$.

In Fig.~\ref{fig:fig2} we plot variations in time of the QFIs calculated analytically based on (\ref{eq:evolution_spin_basis}) for the representative interferometers $\hat{\Lambda}^{(I)}_{\mathbf{n}}= \hat{Q}_{yz}$, $\hat{\Lambda}^{(II)}_{\mathbf{n}}=\hat{D}_{xy}$ and $\hat{\Lambda}^{(III)}_{\mathbf{n}}=\hat{Y}$. In addition $\hat{\Lambda}^{(I)}_{\mathbf{n}}=\hat{J}_x(\gamma)$, with $\gamma$ chosen such that it maximizes the QFI value, is also shown for comparison.
All of them demonstrate Heisenberg-like scaling of the QFI. 
It is important to stress that the QFI value starts from $4 N$ when the interferometer is $\hat{\Lambda}^{(I)}_{\mathbf{n}} =\hat{Q}_{yz}$, and from $0$ for the remaining two  $\hat{\Lambda}^{(II)}_{\mathbf{n}} = \hat{D}_{xy}$ and $\hat{\Lambda}^{(III)}_{\mathbf{n}} = \hat{Y}$. It means that $\hat{\Lambda}^{(I)}_{\mathbf{n}}$ is the most optimal for the experimentally relevant situations where $\bar{t}\sim 1/\sqrt{N}$, at least in the ideal case considered. Time evolution of all QFIs is known analytically, as corresponding covariance matrix elements can be expressed in terms of $\langle \hat{N}_{0} \rangle $, $\langle \hat{N}_{0}^2 \rangle $ and $\langle \hat{a}^\dagger_0 \hat{a}^\dagger_0 \hat{a}_1 \hat{a}_{-1} \rangle$ whose time evolution was presented in the previous section. 
Notice, other choices of interferometers composed of a linear superposition of SU(3) algebra generators are also possible, but their usage will lead to lower values of the QFI than ones obtained from the three optimal interferometers established by us.
Alternatively, the very initial time evolution can be treated under the undepleted pump approximation~\cite{PhysRevLett.117.143004,PhysRevLett.117.013001,PhysRevLett.115.163002,PhysRevA.77.023616} in which the macroscopically populated mode $m_F=0$ acts as a source and injects atoms to the side modes. However, the approximation overestimates the values of the QFI in later times as discussed and demonstrated in \ref{app:undepleted}.

\begin{figure}[]
	\centering
  \includegraphics[width=0.8\linewidth]{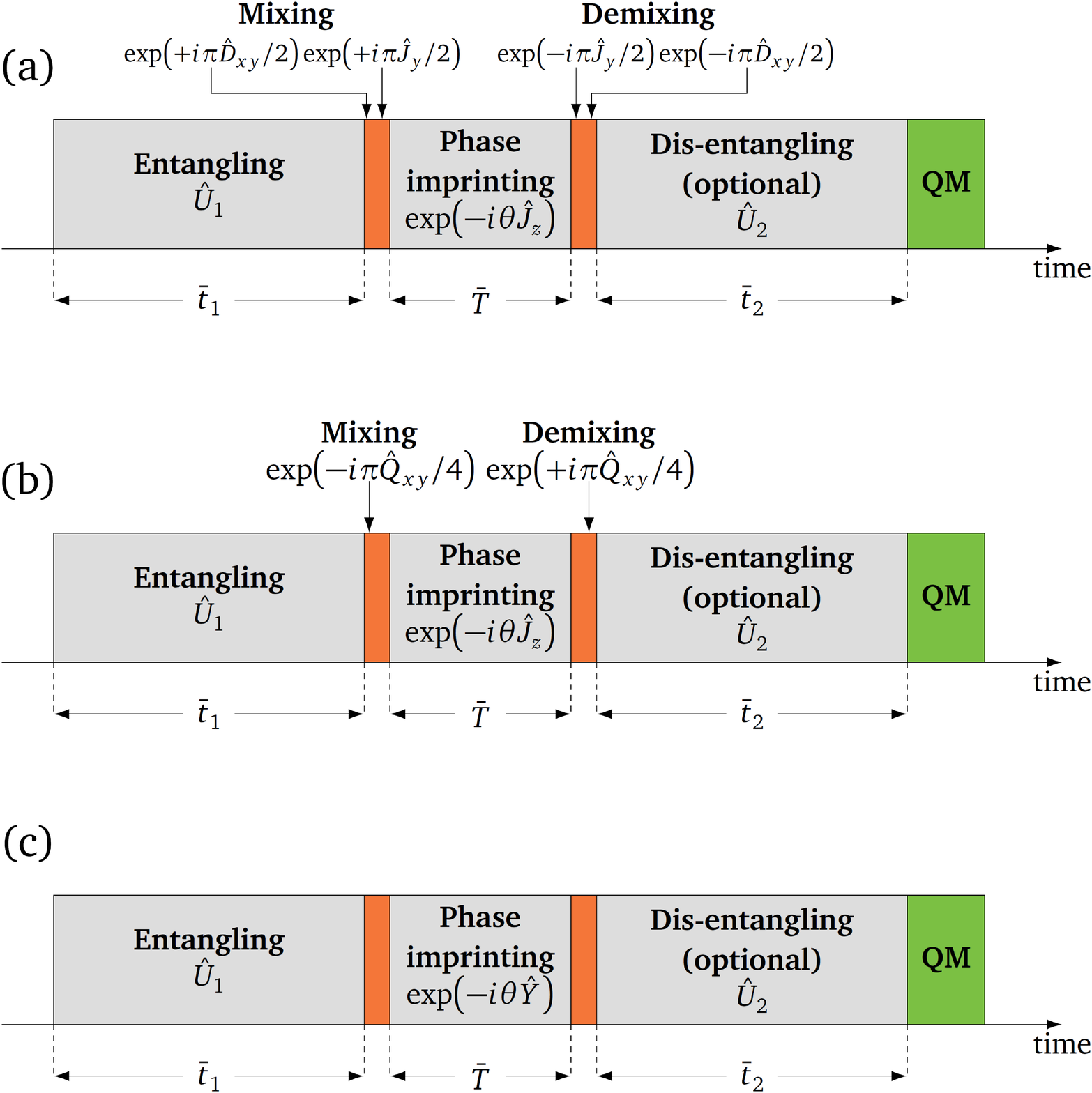}
 \caption{A general protocol for the optimal linear entanglement-enhanced quantum interferometry with spinor condensates. The optimal interferometric rotations discussed in the text are (a) $\hat{\Lambda}^{(I)}_{\mathbf{n}}= \hat{Q}_{yz} $, (b) $\hat{\Lambda}^{(II)}_{\mathbf{n}}= \hat{D}_{xy}$ and (c) $\hat{\Lambda}^{(III)}_{\mathbf{n}}= \hat{Y}$.}
   
    \label{fig:fig3}
\end{figure}

The best interferometric configurations identified in this section are summarized in Fig.~\ref{fig:fig3}. They are quite abstract at the moment, however one can associate them with a measurement of physical quantities such as e.g. magnetic field. Let us consider atomic magnetometers based on detection of the Larmor frequency $\omega$ induced by a weak magnetic field oriented along the $z$ axis. During the Larmor precession cycle the state $\hat{\rho}$ is subject to the phase imprinting process and is effectively rotated around the operator $\hat{J}_z$:
$\hat{\rho}(\theta)=e^{- i \theta \hat{J}_z} \hat{\rho}e^{ i \theta \hat{J}_z}$
with $\theta = \omega \bar{T}$. One can employ $\hat{\Lambda}^{(I)}_{\mathbf{n}}$ and $\hat{\Lambda}^{(II)}_{\mathbf{n}}$ interferometers in that physical situation by a three stage procedure~\cite{KajtochPhd2018}. In order to realize a general rotation one needs to find a unitary transformation $\hat{R}$ such that
\begin{equation}
    \hat{R}^\dagger e^{-i \theta \hat{J}_z} \hat{R} = e^{-i \theta\hat{\Lambda}_{\mathbf{n}} }.
\end{equation}
The procedure is as follows: (i) after the preparation time $\bar{t}=\bar{t}_1$ the state is rotated, resulting in $\hat{\rho}_R=\hat{R}^\dagger \rho \hat{R}$, (ii) the rotated state $\hat{\rho}_R$ is subject to the phase imprinting process $\hat{\rho}_R(\theta)=e^{- i \theta \hat{J}_z} \hat{\rho}_Re^{ i \theta \hat{J}_z}$ and (iii) the state is dis-rotated using the conjugate rotation $\hat{R}^\dagger$ giving $\hat{\rho}(\theta)=\hat{R}^\dagger \hat{\rho}_R(\theta) \hat{R}$. It is straightforward to show that for the first interferometric rotation with $\hat{\Lambda}^{(I)}_{\mathbf{n}}=\hat{Q}_{yz}$ one has the unitary transformation 
$\hat{R}^{(I)}=e^{-i\pi \hat{J}_y/2}e^{-i\pi \hat{D}_{xy}/2}$,
while for $\hat{\Lambda}^{(II)}_{\mathbf{n}}=\hat{D}_{xy}$ one can find that  $\hat{R}^{(II)}=e^{-i\pi \hat{Q}_{xy}/4}$. Unitary transformations $\hat{R}^{(I)}$ and $\hat{R}^{(II)}$ may be realized experimentally as they involve either spin operators or two extremal modes $m_F=\pm 1$ \cite{Lucke2011, PhysRevA.88.031602}. 
In the case of the third optimal interferometer found by us, namely $\hat{\Lambda}^{(III)}_{\mathbf{n}}=\hat{Y}$, the physical interpretation is already understood very well within $SU(1,1)$ interferometry and was realized experimentally in \cite{OberthalerSU11}.

%%%%%%%%%%%%%%%%%%%%%%%%%%%%%%%%%%%%%%%%%%%%%%%%%%%%%%%%%%%%%%%%%%%%%%%%%%%%%%%%%%%%%%%%%%%%%%%%%
\section{Identification of optimal observables}\label{sec:observable}

\begin{figure}[]
\centering
\includegraphics[width=0.56\linewidth]{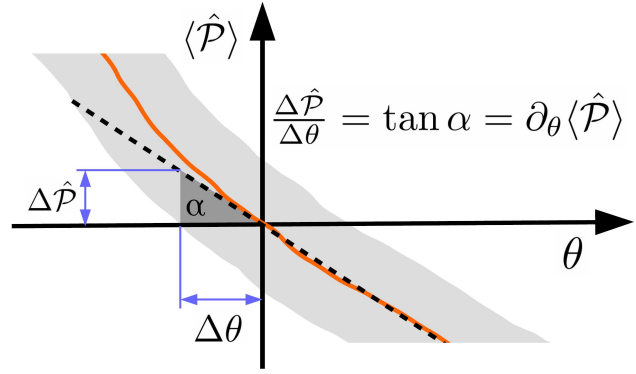}
\caption{The origin of the error-propagation formula (\ref{eq:errorpropagation}). The precision in the $\theta$ estimation is based on the measurement of a signal $\langle \hat{\mathcal{P}} \rangle $ which is illustrated by the orange solid curve. The uncertainty of $\hat{\mathcal{P}}$ is marked by the gray shadow region. The tangent of the curve $\langle \hat{\mathcal{P}} \rangle $ at some value of $\theta$ can be determined by its slope ${{\partial }_{\theta }}\langle \hat{\mathcal{P}} \rangle $ and by the ratio $\Delta \hat{\mathcal{P}}/\Delta \theta $. Their equivalence leads to the error-propagation formula (\ref{eq:errorpropagation}).}
\label{fig:fig4}
\end{figure}

In the quantum interferometry scheme, a physical quantity like magnetic field is mapped onto the phase difference $\theta$ between internal states of atoms. Then it can be extracted by performing a quantum measurement.
The expectation value of the observable $\langle \hat{\mathcal{P}} \rangle $ carries information about the unknown value of $\theta$, and thus can be exploited in the estimation procedure. 
As illustrated in Fig.~\ref{fig:fig4}, in the limit of a large number of measurements the precision in the $\theta$ estimation is given by the error-propagation formula~\cite{1751-8121-47-42-424006,RevModPhys.90.035005,Smerzi}: 
\begin{equation}\label{eq:errorpropagation}
 \Delta ^{-2} \theta (\hat{\mathcal{P}}) = \frac{| \partial_\theta \langle \hat{\mathcal{P}} \rangle|^2}{\Delta^2 \hat{\mathcal{P}}},
\end{equation}
where $\Delta ^2 \hat{\mathcal{P}} = \langle \hat{\mathcal{P}}^2 \rangle - \langle \hat{\mathcal{P}}\rangle^2$. 
We have already mentioned in the previous section that the uncertainty of $\theta$ is bounded from below by the QFI, namely $ \Delta^{-2} \theta(\hat{\mathcal{P}}) \leq F_Q$, which is nothing else but the Cram\`er-Rao inequality. 

In principle, it is possible to choose such an observable $\langle \hat{\mathcal{P}} \rangle$ which saturates the Cramer-Rao inequality.
The natural question arises which $\langle \hat{\mathcal{P}} \rangle$ provides the highest precision under the three optimal interferometric rotations we identified in the previous section, i.e. $\hat{Q}_{yz}, \hat{D}_{xy},  \hat{Y}$~?

When identifying such observables, it is important to take into account detection imperfections. 
Entangled states which are generated in time with the Hamiltonian (\ref{eq:ham}) provide limited sensitivity due to the requirement of perfect detection, sometimes on the level of a single particle. 
In the nowadays experiments, the measurement is burdened with the particle detection noise, therefore imperfect detection of a state causes a significant drop of the QFI value. 
In a standard ultra-cold atom setup information about various physical quantities is estimated from the measurement of the atom number. In an ideal system, the probability $p(N|\bar{N})$ of detecting $N$ number of atoms given that the true number is $\bar{N}$ equals $\delta_{N,\bar{N}}$. In a realistic scenario, that property no longer holds and one can detect $N$ number of atoms even though $\bar{N} \neq N$ truly hit the detector. 

Mathematically, detection noise is modeled by replacing all ideal probabilities $p(\bar{N}|\theta)$ with $\tilde{p}(N|\theta) = \sum_{\bar{N}}p(N|\bar{N}) p(\bar{N}|\theta)$, where $\sum_N p(N|\bar{N}) = 1$~\cite{PhysRevD.33.1643,PhysRevLett.110.163604}. It can be shown by simple algebra that the $k$th moment of the particle number operator is modified by the Gaussian detection noise $p(N|\bar{N}) = \exp[-(N-\bar{N})^2/2\sigma^2]/(\sigma \sqrt{2\pi})$ in the following way:
\begin{equation}
\langle\hat{N}^{k} \rangle_{\rm gdn} = \sum\limits_{N} N^{k} \tilde{p}(N|\theta)  \simeq \sum\limits_{l=0}^{k}\binom{k}{l} M_l(\sigma) \langle \hat{N}^{k-l}\rangle_{\rm id}, 
\label{eq:detection_noise_derivation}
\end{equation}
where $M_l(\sigma)=\int x^l {\rm exp}(-x^2/2\sigma^2)/(\sigma\sqrt{2\pi})dx$ is the $l$th central moment of the Normal distribution and $\langle \hat{N}^{k-l}\rangle_{\rm id}=\sum_{\bar{N}} \bar{N}^{k-l}p(\bar{N},\theta)$ is the ideal expectation value without detection noise, e.g. $\langle \hat{N}^2 \rangle_{\rm gdn} = \langle \hat{N}^2 \rangle_{\rm id} + \sigma^2$. In Eq.~\eqref{eq:detection_noise_derivation} we assumed that the atom number is large enough so that the difference between the sum $\sum_N N^k p(N|\bar{N})$ and the integral $\int dN\ N^k p(N|\bar{N})$ is negligible. Notice, the effect of the Gaussian detection noise on the moments of operator $\hat{N}$ is the same as if it was replaced by $\hat{N} = \hat{N}^{0} + \delta\hat{N}$, where $\hat{N}^{0}$ denotes an ideal particle number operator and $\delta \hat{N}$ is an independent random operator satisfying $\langle \delta \hat{N} \rangle = 0$ and $\langle (\delta \hat{N})^2\rangle = \sigma^2$~\cite{PhysRevA.73.013814}.

The Gaussian detection noise does not modify the derivative entering in the error-propagation formula (\ref{eq:errorpropagation}) but adds to the variance typically as follows:
\begin{equation}\label{eq:gdn_errorpropagation}
\Delta^{-2}\theta_{\rm gdn}(\hat{\mathcal{P}}) = \frac{|\partial_{\theta}\langle \hat{\mathcal{P}}\rangle|^2}{\Delta^2 \hat{\mathcal{P}}_{\rm gdn} + \sigma^{2}_{\mathcal{P}}},
\end{equation}
where $\sigma^{2}_{\mathcal{P}}$ is connected with the second moment of $\hat{\mathcal{P}}$.
In the case of our model we  assume that the width of the Gaussian distribution $\sigma$ is the same for the probabilities of detecting atoms in all three Zeeman components.

The interaction-based readout, very nicely introduced and explained e.g. in \cite{PhysRevLett.116.053601,PhysRevLett.118.150401,PhysRevLett.119.193601,PhysRevA.94.010102}, helps to avoid direct detection of entangled states, and therefore, protects against the noise effect. Typically, it is done by the time-reversed evolution and we will start our analysis with this protocol. In the further part of this work we will use these methods while recognizing the observable $\langle \hat{\mathcal{P}} \rangle$ which saturates the Cramer-Rao inequality under the optimal interferometric rotations determined by us in the previous section.

%%%%%%%%%%%%%%%%%%%%%%%%%%%%%%%%%%%%%%%%%%%%%%%%%%%%%%%%%%%%%%%%%%%%%%%%%%%%%%%%%%%%%%%%%%%%%%%%%
\subsection{Interaction-based readout to protect against the detection noise}

The interaction-based readout protocol is simply a unitary evolution based on the time reversed non-linear interactions applied after the phase imprinting operation~\cite{PhysRevLett.116.053601, doi:10.1117/12.2257033}, and is defined in the following way:
\begin{equation}\label{eq:evol_detec}
|\psi(\theta)\rangle_{\rm id} = \hat{U}_2 e^{-i\theta \hat{\Lambda}_{\mathbf{n}}}\hat{U}_1|0,N,0\rangle .
\end{equation}
The protocol considered by us fits to the one sketched in Fig.~\ref{fig:fig1}, and contains non-trivial unitary operation $\hat{U}_2=e^{i \bar{t}_2 \hat{J}^2}$, with $\bar{t}_2=\bar{t}_1$ for simplicity. 

In the case of spin-1 condensates the interaction based readout was already experimentally realized in the context of SU(1,1) interferometry \cite{PhysRevLett.117.013001, OberthalerSU11}
and theoretically analyzed \cite{PhysRevA.97.043813, PhysRevLett.119.193601}. In the latter, it is the Fisher information~\footnote{The general definition of the classical Fisher Information~\cite{braunstein1994} is given by $\mathcal{I}(\theta) = \sum_x p(x|\theta)^{-1}(\partial_\theta p(x|\theta))^2$, where $p(x|\theta)=Tr(\hat{\Pi}_x\hat{\rho}_{out})$ is a conditional probability of measuring the outcome $x$ with given $\theta$, while $\hat{\Pi}_x$ is the measurement operator satisfying $\sum_x \hat{\Pi}_x^\dagger \hat{\Pi}_x = \mathbf{1}$. 
The knowledge of $p(x|\theta)$ is used to
construct an estimator for the phase $\theta$, and according to
the Cram\'er-Rao inequality, the precision $\Delta\theta$ in the $\theta$ estimation is bounded from below by the Fisher information $
\mathcal{I}(\theta)$.} used to quantify the Cram\`er-Rao inequality. In our paper we use instead the error propagation formula (\ref{eq:gdn_errorpropagation}) and look for the optimal observables that saturate the inequality in the experimentally relevant short times range.
It turns out that the one among realatively easy to measure experimentally observables, namely $\hat{\mathcal{P}} =  \hat{J}_z^2 , \, \hat{Y}, \,  \hat{Q}_{yz}$, is sufficient to saturate the corresponding QFI value with particular choice of the generator of the interferometric rotation. 

%Qyz
\subsubsection{The first optimal interferometer $\hat{\Lambda}_{\bf n}^{(I)}=\hat{Q}_{yz}$}

\begin{figure}[]
\centering
\begin{picture}(0,170)
\put(-130,-5){\includegraphics[width=0.6\linewidth]{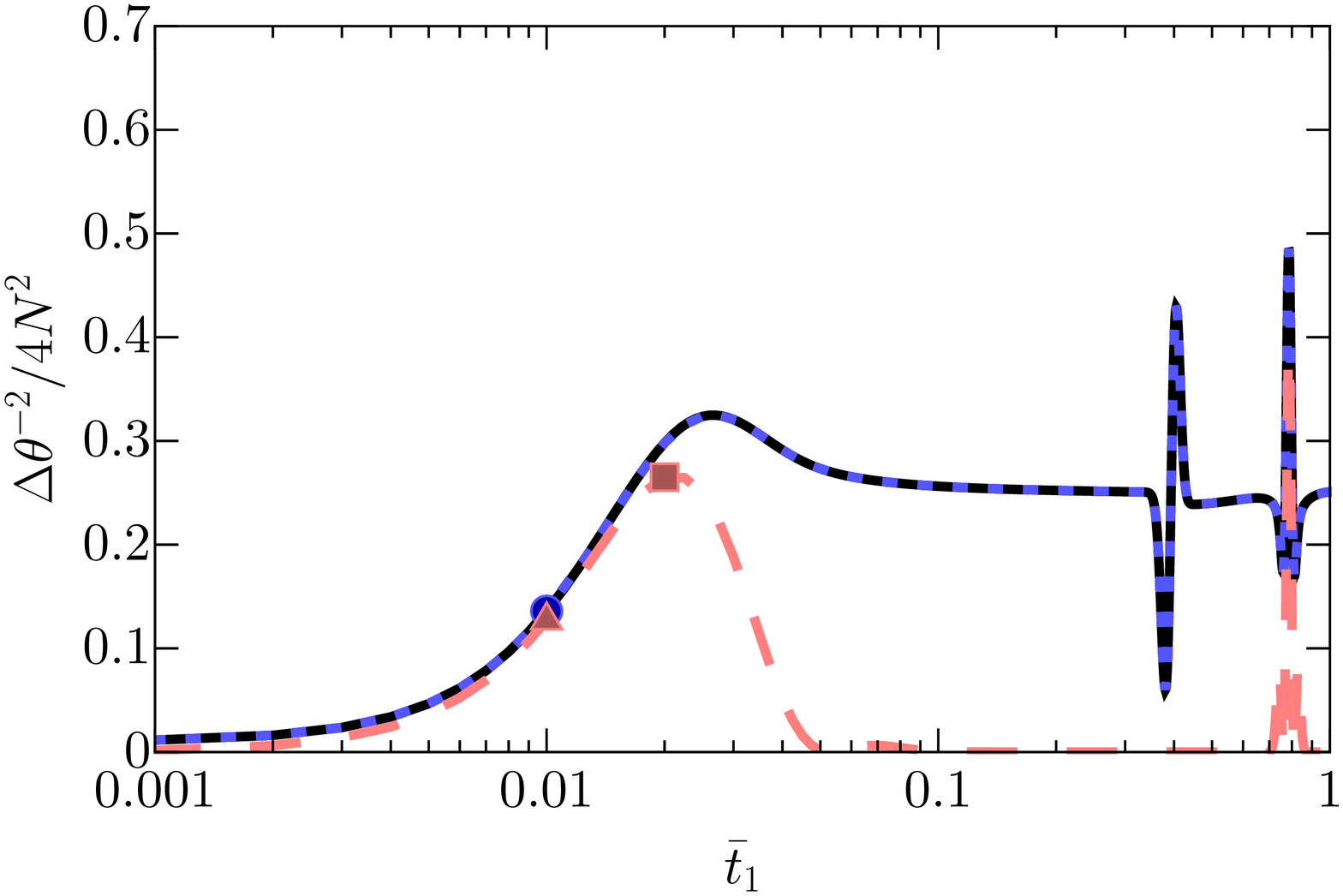}}
\put(-95,82){\includegraphics[width=0.27\linewidth]{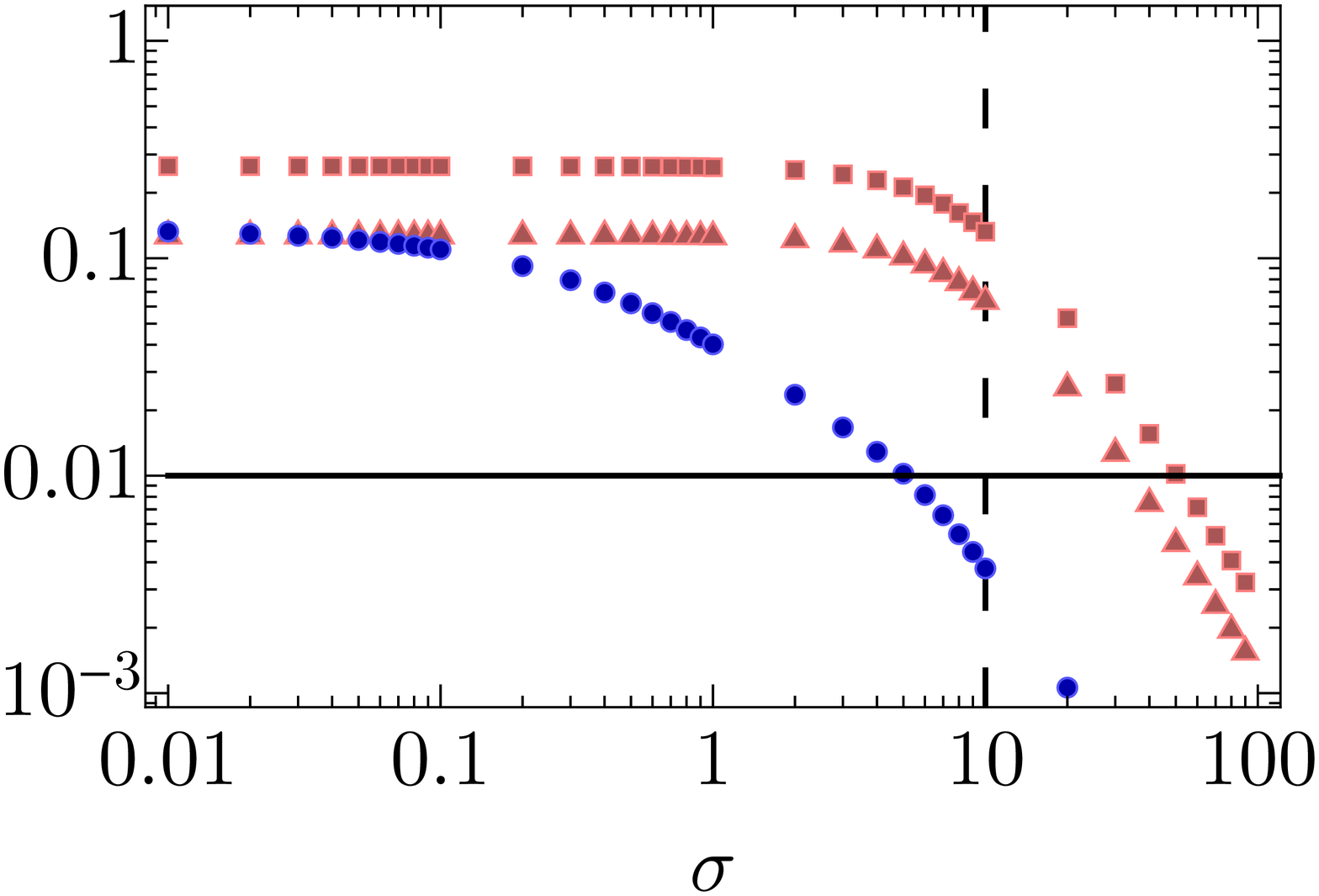}}
\end{picture}
\caption{The inverse of the precision in the $\theta$ estimation from the error-propagation formula (\ref{eq:QyzQFI}) for $\hat{\mathcal{P}}=\hat{J}_z^2$ (blue dotted line) and from \eqref{eq:QyzQYZQFI} for $\hat{\mathcal{P}}=\hat{Q}_{yz}$ (pink dashed line), both with the first optimal interferometer $\hat{\Lambda}_{\bf n}^{(I)}=\hat{Q}_{yz}$. The value of $F_Q^{(I)}$ (black solid line) is shown for comparison. 
The inset shows variations of $\Delta^{-2} \theta^{(I)}_{\rm gdn}(\hat{J}_z^2)$ (blue circles) and $\Delta^{-2} \theta^{(I)}_{\rm gdn}(\hat{Q}_{yz})$ (pink triangles and squares) versus $\sigma$ for $\bar{t}_1=0.01$ and $\bar{t}_1=0.02$ as indicated by the corresponding points in the main plot. Here the black solid line marks the position of the SQL and the black dashed vertical line stands for $\sigma=\sqrt{N}$.}
\label{fig:fig5}
\end{figure}

When the first optimal interferometer $\hat{\Lambda}_{\bf n}^{(I)}$ is used in the protocol  (\ref{eq:evol_detec}), then the inverse of the uncertainty calculated from the error-propagation formula (\ref{eq:errorpropagation}) for $ \hat{\mathcal{P}} =\hat{J}_z^2 $, i.e. 
\begin{equation}\label{eq:QyzQFI}
\Delta^{-2} \theta^{(I)}_{\rm gdn}(\hat{J}_z^2) = \frac{| \partial_\theta \langle \hat{J}_z^2 \rangle_{\rm id}|^2}
{\Delta^2 ( \hat{J}_z^2)_{\rm id} + 4( \hat{J}_z^2)_{\rm id}\sigma^2 +  2\sigma^4} ,
\end{equation}
saturates the Cram\`er-Rao inequality as long as $\sigma\to 0$, what is demonstrated in Fig.~\ref{fig:fig5}. That fantastic agreement should be possible to prove analytically, however we were not succeeded. Notice, the right hand side of Eq.(\ref{eq:QyzQFI}) is $0/0$ expression when $\theta \to 0$ and $\sigma\to 0$ 
because $\partial_\theta \langle \hat{J}_z^2 \rangle_{\rm id}\to \langle \comm{\hat{Q}_{yz}}{\hat{J_z^2}} \rangle \propto \langle \hat{Q}_{zx}\hat{J_z} + \hat{J_z} \hat{Q}_{zx} \rangle \to 0$ and $\Delta^2 ( \hat{J}_z^2)_{\rm id} \to 0$ in the initial state. A direct consequence that $\hat{J}^2_z$ commutes with both $\hat{U}_{1,2}$ is a fast drop of the $F_Q^{(I)}$ value when $\sigma$ increases. It means that the detection noise reduces the signal's value, although it is so simple to measure. 

One can overcome the problem and measure $ \hat{\mathcal{P}} =\hat{Q}_{yz}$ in place of $\hat{J}_z^2$. The measurement of $\langle \hat{Q}_{yz} \rangle$ is possible using nowadays technology with the appropriate choice of state rotations which map the value of $\langle \hat{Q}_{yz} \rangle$ onto the value of $\langle \hat{J}_{z} \rangle$~\cite{Hamley2012}. Then, as can be seen in Fig.~\ref{fig:fig5}, the short time dynamics achievable in nowadays experiments is quite well captured. The disadvantage of that choice of $\hat{\mathcal{P}}$ is the zero value of $\Delta^{-2} \theta^{(I)}_{\rm gdn}(\hat{Q}_{yz}) $ when $t\to 0 $ and $\sigma\to 0$, so one does not start from SQL as it was the case for $ \hat{\mathcal{P}} =\hat{J}_{z}^2$. The resulting inverse of the uncertainty is insensitive to the detection noise from the error-propagation formula \eqref{eq:errorpropagation} 
\begin{equation}\label{eq:QyzQYZQFI}
\Delta^{-2} \theta^{(I)}_{\rm gdn}(\hat{Q}_{yz}) = \frac{| \partial_\theta \langle \hat{Q}_{yz} \rangle_{\rm id}|^2}
{\Delta^2 ( \hat{Q}_{yz})_{\rm id}  +  \sigma^2} ,
\end{equation}
which is demonstrated in the inset of Fig.~\ref{fig:fig5}. Finally, the first maximum of $\Delta^{-2} \theta^{(I)}_{\rm gdn}(\hat{Q}_{yz})$ at $\theta \to 0$ has the Heisenberg scaling, see Fig.~\ref{fig:fig5}. This configuration seems to be the most promising for quantum metrology beyond the SQL and it was not realized experimentally yet, although it is in the range of the present technology.

%Dxy
\subsubsection{The second interferometer $\hat{\Lambda}_{\bf n}^{(II)}=\hat{D}_{xy}$}

\begin{figure}[]
\centering
\begin{picture}(0,170)
\put(-130,-5){\includegraphics[width=0.6\linewidth]{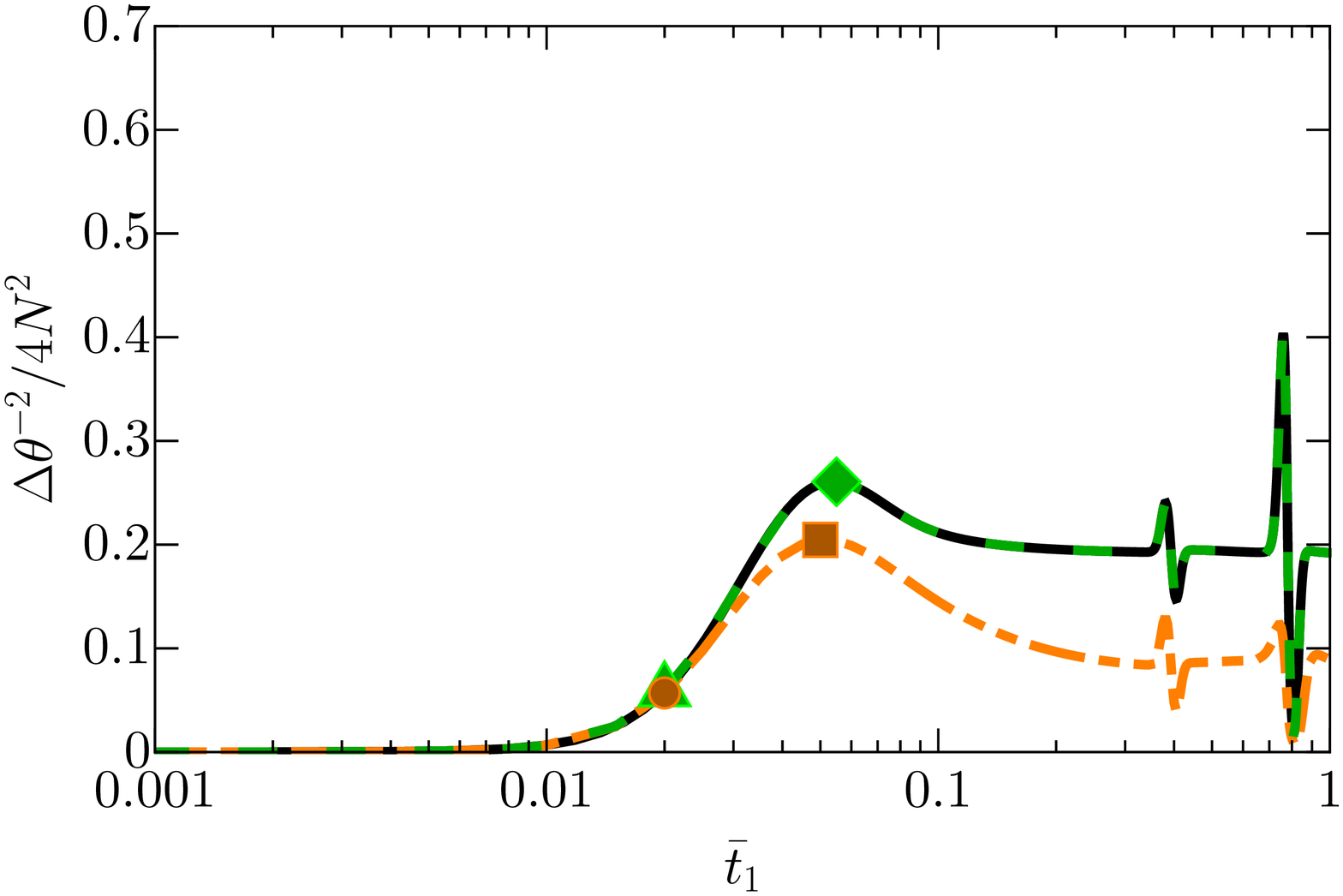}}
\put(-95,70){\includegraphics[width=.27\linewidth]{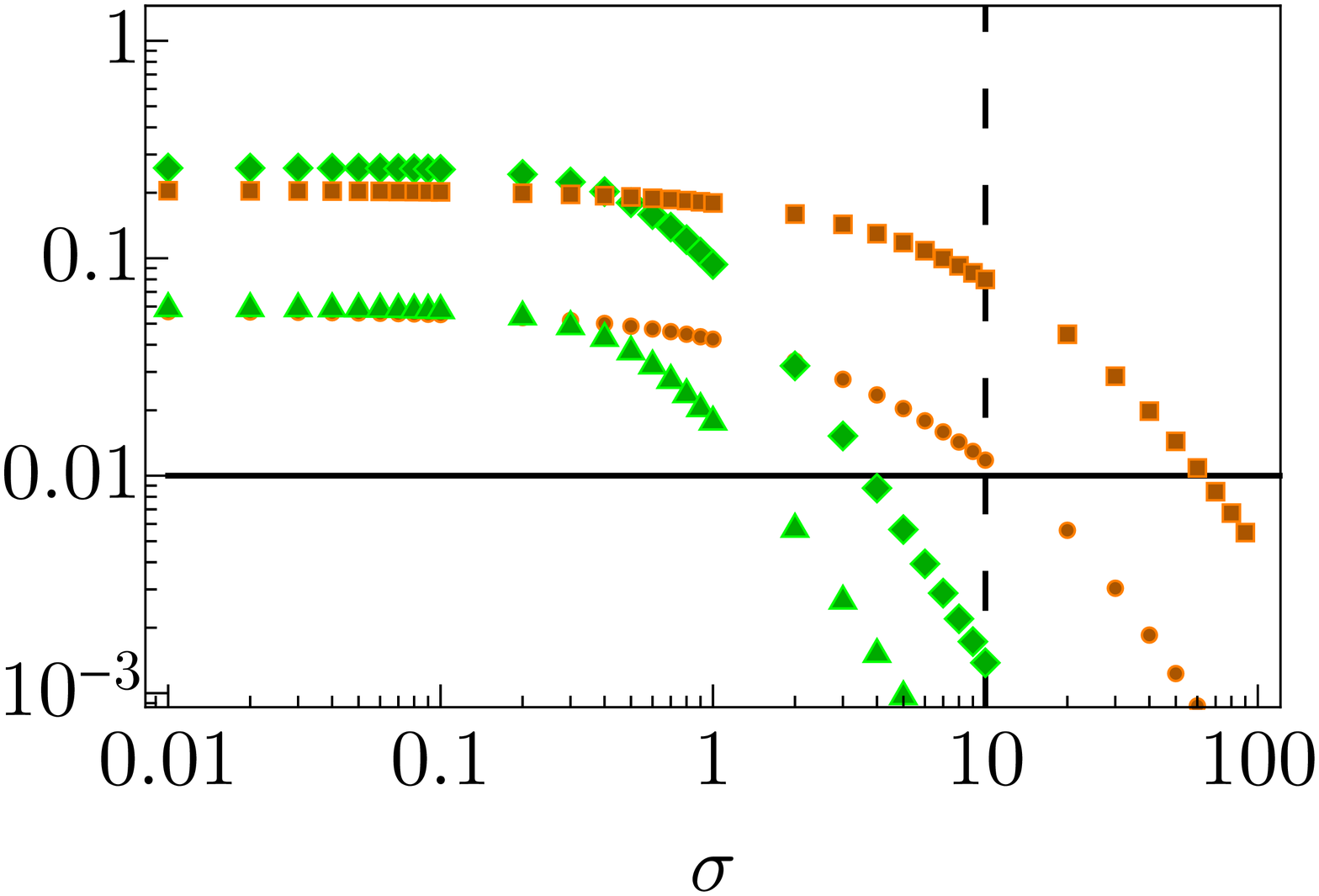}}
\end{picture}
\caption{The inverse of the precision in the $\theta$ estimation from the error-propagation formula (\ref{eq:DxyQFI}) for $\hat{\mathcal{P}}=\hat{J}_z^2$ (green dashed line) and $\hat{\mathcal{P}}=\hat{Y}$ (orange dash double dotted line), both for $\hat{\Lambda}_{\bf n}^{(II)}=\hat{D}_{xy}$ compared with $F_Q^{(II)}$ (black solid line). 
The inset shows $\Delta^{-2} \theta^{(II)}_{\rm gdn}(\hat{J}_z^2)$ as a function of the detection noise $\sigma$ for $\bar{t}_1=0.02$ (green triangles) and $\bar{t}_1=0.055$ (green diamonds) and $\Delta^{-2} \theta^{(II)}_{\rm gdn}(\hat{Y})$ versus $\sigma$ for $\bar{t}_1=0.02$ (orange circles) and $\bar{t}_1=0.05$ (orange squares). 
Here, the black solid line marks the position of the SQL and the black dashed vertical line stands for $\sigma=\sqrt{N}$.}
\label{fig:fig6}
\end{figure}

The second among optimal interferometers, namely $\hat{\Lambda}_{\bf n}^{II}= \hat{D}_{xy}$, was already realized experimentally ~\cite{Lucke2011} and the measurement of $\hat{J}_z^2$ was performed. Here, we just demonstrate the origin and rightness of such a choice pointing out its sensitivity to the detection noise. Indeed, one can show numerically that
\begin{equation}\label{eq:DxyQFI}
\Delta^{-2} \theta^{(II)}_{\rm gdn}(\hat{J}_z^2) = \frac{| \partial_\theta \langle \hat{J}_z^2 \rangle_{\rm id}|^2}
{\Delta^2 ( \hat{J}_z^2)_{\rm id} + 4( \hat{J}_z^2)_{\rm id}\sigma^2 +  2\sigma^4}
\end{equation}
saturates the Cram\`er-Rao inequality, as $\Delta^{-2} \theta^{(II)}_{\rm gdn}(\hat{J}_z^2)\simeq F_Q^{(II)}$ whenever $\sigma \to 0$, as demonstrated in Fig.~\ref{fig:fig6} by the green dashed line. However, the operator $\hat{\mathcal{P}}=\hat{J}^2_z$ commutes with both $\hat{U}_{1}$ and $\hat{U}_{2}$, which makes the precision very sensitive to the detection noise. The non-zero value of $\sigma$ decreases the value of $\Delta^{-2} \theta^{(II)}_{\rm gdn}(\hat{J}_z^2)$ as shown in the inset of Fig.\ref{fig:fig6}. Nevertheless, we found out that there is another signal that do not share this property and it is as simple to measure experimentally as the previous one, namely $\mathcal{\hat{P}}=\hat{Y}$. The inverse of the uncertainty calculated from the error-propagation formula \eqref{eq:errorpropagation} for this signal is given by
\begin{equation}\label{eq:DXYYQFI}
\Delta^{-2} \theta^{(II)}_{\rm gdn}(\hat{Y}) = \frac{| \partial_\theta \langle \hat{Y} \rangle_{\rm id}|^2}{\Delta^2 ( \hat{Y})_{\rm id} + \sigma^2}.
\end{equation}
Indeed, in the very initial period of time we achieve the same result as in the case of $\mathcal{\hat{P}}=\hat{J}_z^2$. Although, proceeding with the evolution little further causes a drop of the Fisher information. Nevertheless, initially the value of $\Delta^{-2} \theta^{(II)}_{\rm gdn}(\hat{Y})$ gains resistance to the detection noise as presented in the inset of Fig.~\ref{fig:fig6}.

%Y
\subsubsection{When $\hat{\Lambda}_{\bf n}^{(III)}=\hat{Y}$}

\begin{figure}[]
\centering
\begin{picture}(0,170)
\put(-130,-5){\includegraphics[width=0.6\linewidth]{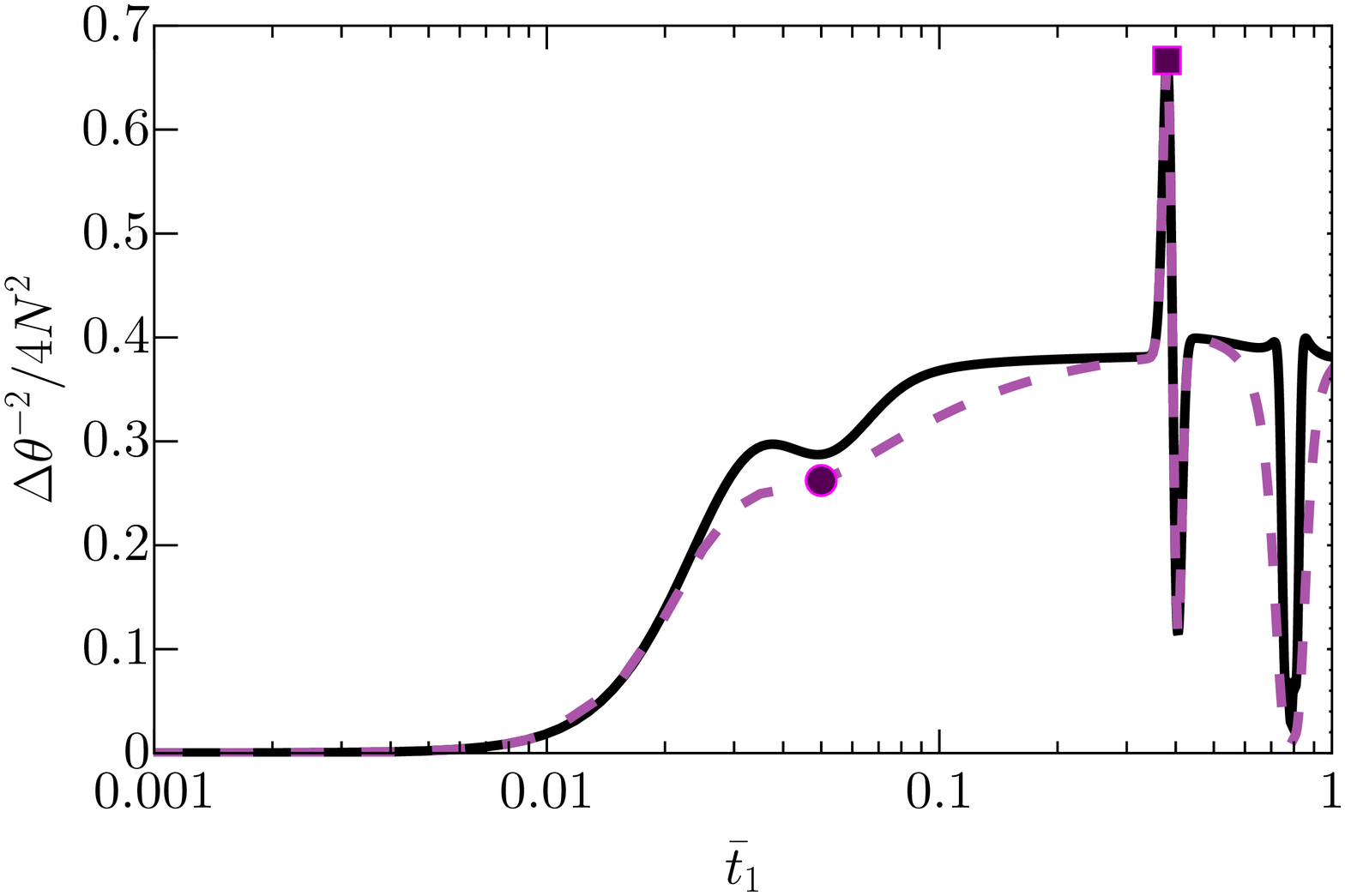}}
\put(-95,78){\includegraphics[width=.27\linewidth]{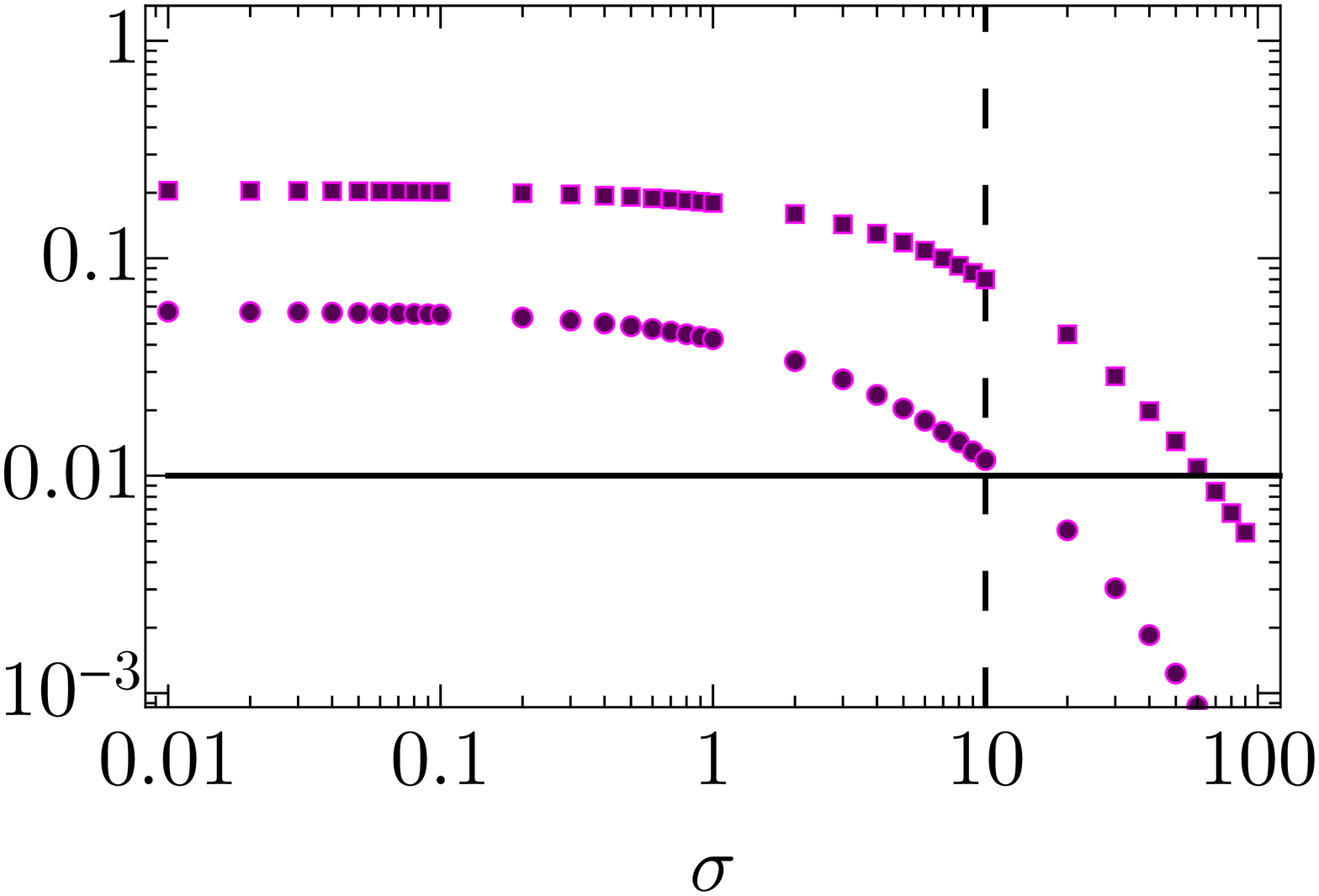}}
\end{picture}
\caption{The inverse of the precision in the $\theta$ estimation from the error-propagation formula (\ref{eq:YQFI}) for $\hat{\mathcal{P}}=\hat{Y}$ (purple dashed line) with $\hat{\Lambda}_{\bf n}^{(III)}=\hat{Y}$ compared to the value of the quantum Fisher information $F_Q^{(I)}$ (black solid line). The inset shows $\Delta^{-2} \theta^{(I)}_{\rm gdn}(\hat{Y})$ as a function of the detection noise $\sigma$ for $\bar{t}_1=0.05$ (purple circles) and $\bar{t}_1=0.38$ (purple squares). 
The black solid line marks the position of the SQL and the black dashed vertical line stands for $\sigma=\sqrt{N}$.}
\label{fig:fig7}
\end{figure}

Finally, the third interferometer identified by us $\hat{\Lambda}_{\bf n}= \hat{Y}$ was realized experimentally~\cite{PhysRevLett.117.013001, OberthalerSU11} as well, however in the context of the $SU(1,1)$ interferometry. The measurement of $\hat{Y}$ was performed. Indeed, our calculations confirm that it is an optimal choice, as
\begin{equation}\label{eq:YQFI}
\Delta^{-2} \theta^{(III)}_{\rm gdn}(\hat{Y}) = \frac{| \partial_\theta \langle \hat{Y} \rangle_{\rm id}|^2}{\Delta^2 ( \hat{Y})_{\rm id} + \sigma^2}
\end{equation}
saturates the Cram\`er-Rao inequality, i.e $\Delta^{-2} \theta^{(III)}_{\rm gdn}(\hat{Y})\simeq F_Q^{(III)}$, whenever $\sigma \to 0$ as illustrated in Fig.~\ref{fig:fig7}. Moreover, our calculations show that the uncertainty from the error propagation (\ref{eq:YQFI}) formula is insensitive to detection noise.

In summary, for all of the three interferometers we recognized in the previous section it is possible to choose quite simple to measure experimentally signals $\mathcal{\hat{P}}$ in such a way that they are insensitive to the detection noise, preventing a drop of the signal's value. In addition to the known in the literature configurations, we found out the additional one which gives desired precision much faster in time than the remaining two.

%%%%%%%%%%%%%%%%%%%%%%%%%%%%%%%%%%%%%%%%%%%%%%%%%%%%%%%%%%%%%%%%%%%%%%%%%%%%%%%%%%%%%%%%%%%%%%%%%
\subsection{Proposal for an experimental realization of interaction-based readout by a single rotation}

\begin{figure*}[]
\centering
{\includegraphics[width=0.99\linewidth]{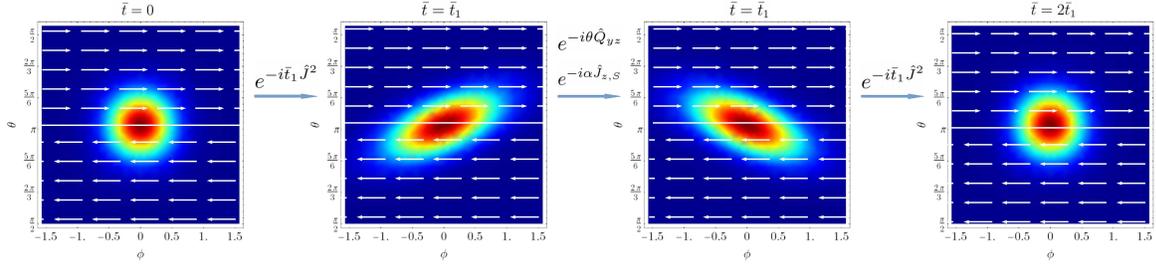}}
\caption{Mercator projections of Husimi functions for evolution given with \eqref{eq:evol_invert} in the symmetric subspace spanned by the three operators $\{\hat{J}_x, \hat{Q}_{yz}, \frac{1}{2}(\sqrt{3}\hat{Y}+\hat{D}_{xy})\}$. Superimposed white lines show mean-field phase portraits explained in the text. The initial coherent state $|0,N,0\rangle$ is located along the $z$ axis of the generalized Bloch sphere and is squeezed at some moment of time $\bar{t}=\bar{t}_1$. Next, the state is imposed to the phase imprinting process and an extra rotation around $\hat{J}_{z,S}$, by the proper angle $\alpha$, is applied in order to artificially inverse a further evolution.}
\label{fig:fig8}
\end{figure*}

Now, we concentrate on the first interferometer $\hat{\Lambda}_{\bf n}^{(I)}=\hat{Q}_{yz}$ and the measurement of $ \hat{\mathcal{P}} =\hat{Q}_{yz}$. Indeed, the configuration is very promising and, in addition, insensitive to the detection noise. However, the main objection for experimental realization would be difficulty of implementation of the time inversion, which is necessary to have $\hat{U}_2=e^{i \hat{J}^2 \bar{t}_2}$. In the following we present the method based on a single rotation of the state which works for short enough times.

In order to clearly present our idea, we start consideration by introducing the symmetric and antisymmetric bosonic operators
$\hat{g}_S=(\hat{a}_1+\hat{a}_{-1})/\sqrt{2}$, $\hat{g}_A=(\hat{a}_1-\hat{a}_{-1})/\sqrt{2}$
and spin operators 
\begin{align}
&\hat{J}_{x,l}=\hat{a}^\dagger_0\hat{g}_l+\hat{a}_0\hat{g}^\dagger_l, \\
&\hat{J}_{y,l}=i(\hat{a}^\dagger_0\hat{g}_l-\hat{a}_0\hat{g}^\dagger_l), \\
&\hat{J}_{z,l}=\hat{g}_l^\dagger \hat{g}_l - \hat{a}^\dagger_0\hat{a}_0,
\end{align}
which are symmetric when $l=S$ and anti-symmetric for $l=A$. 
The above spin operators have cyclic commutation relations, e.g. $[\hat{J}_{x,l}, \hat{J}_{y,l}]=2 i \hat{J}_{z,l}$. 
The Hamiltonian that we used in the previous subsections expressed in terms of the symmetric and anti-symmetric operators reads:
\begin{equation}
    \ham=c_2'(\hat{J}_{x, S}^2 + \hat{J}_{y, A}^2 + \hat{J}_{z}^2).
\end{equation}
The SU(2) subspace spanned by the symmetric spin operators is 
$\{\hat{J}_{x,S},\hat{J}_{y,S},\hat{J}_{z,S}\}=\{\hat{J}_x, \hat{Q}_{yz}, \frac{1}{2}(\sqrt{3}\hat{Y}+\hat{D}_{xy})\}$, and the SU(2) subspace spanned by the anti-symmetric spin operators is $\{\hat{J}_{x,A},\hat{J}_{y,A},\hat{J}_{z,A}\}=\{ \hat{Q}_{zx},\hat{J}_y, \frac{1}{2}(\sqrt{3}\hat{Y}-\hat{D}_{xy})\}$. 
Let us concentrate our attention on the symmetric subspace, as both interferometer and measurement are located in it.

The state, as well as its evolution, can be illustrated on the Bloch sphere spanned by the SU(2) symmetric spin operators with the help of the Husimi function 
\begin{equation}
    Q(\theta_Q , \phi_Q)=\left| \langle \theta_Q, \phi_Q | \Psi(t) \rangle\right|^2,
\end{equation}
where arbitrary spin-coherent state in the symmetric subspace is defined as
\begin{equation}\label{eq:coherentstate}
    \ket{\theta_Q,\phi_Q}=
    \sum_{k,J}\sqrt{\binom{N}{k+j}\binom{k+j}{j}}
    \left(\frac{1}{\sqrt{2}} \sin{\frac{\theta_Q}{2}}e^{i\phi_Q}\right)^j
    \left(\frac{1}{\sqrt{2}}\cos{\frac{\theta_Q}{2}}\right)^{N-j}\ket{k,j,N-k-j}.
\end{equation}
The initial coherent state $\ket{0,N,0}$ is located along the $z$ axis of the Bloch sphere as $\langle \hat{J}_{x, S}\rangle = \langle \hat{J}_{y, S}\rangle = 0$ and $\langle \hat{J}_{z, S}\rangle=-N$. The very initial evolution of the state is identical to the one governed by the one-axis twisting model \cite{PhysRevA.47.5138}. Hence, one can predict approximate quantum evolution following the mean-field phase portrait \cite{PhysRevA.92.023603} which is explained in \ref{app:meanfield} and shown in Fig.~\ref{fig:fig8} by white arrows. The initial coherent state $\ket{0,N,0}$ evolves along circulating trajectories and is squeezed initially. The squeezed state can be rotated around the $\hat{J}_{z,S}$ axis of the Bloch sphere, so the further evolving state will ideally turn back to the initial coherent state after the time $\bar{t}=2\bar{t}_1$. The rotation allows us to perform backward evolution needed for implementation of the interaction-based readout protocol. Our idea is explained in details in Fig.~\ref{fig:fig8}.

\begin{figure}[]
\centering
\begin{picture}(0,240)
% a
\put(-160,120)
{\includegraphics[width=0.3696\linewidth]{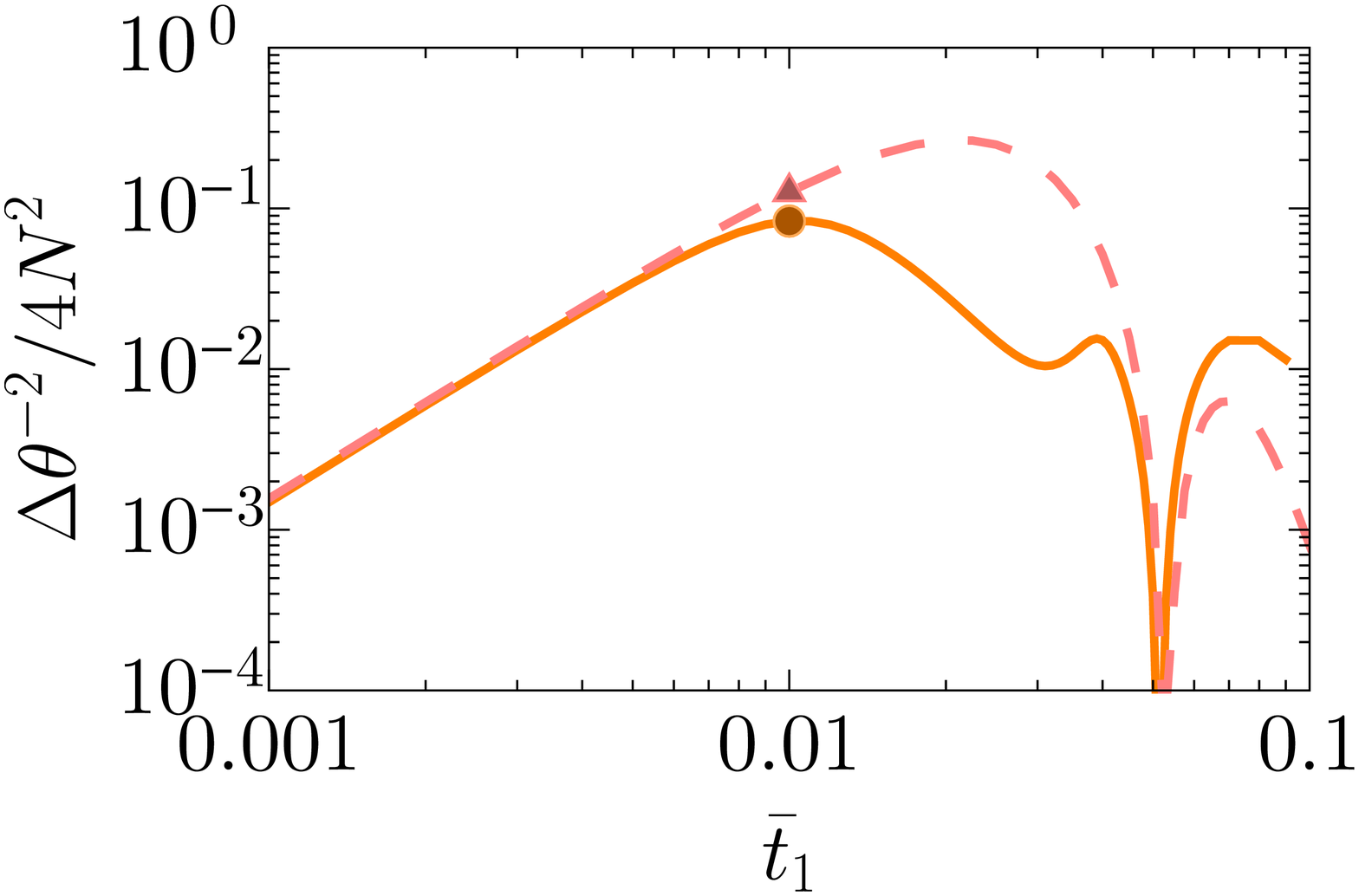}}
\put(-120,210){(a)}
% b
\put(25,120)
{\includegraphics[width=0.3696\linewidth]{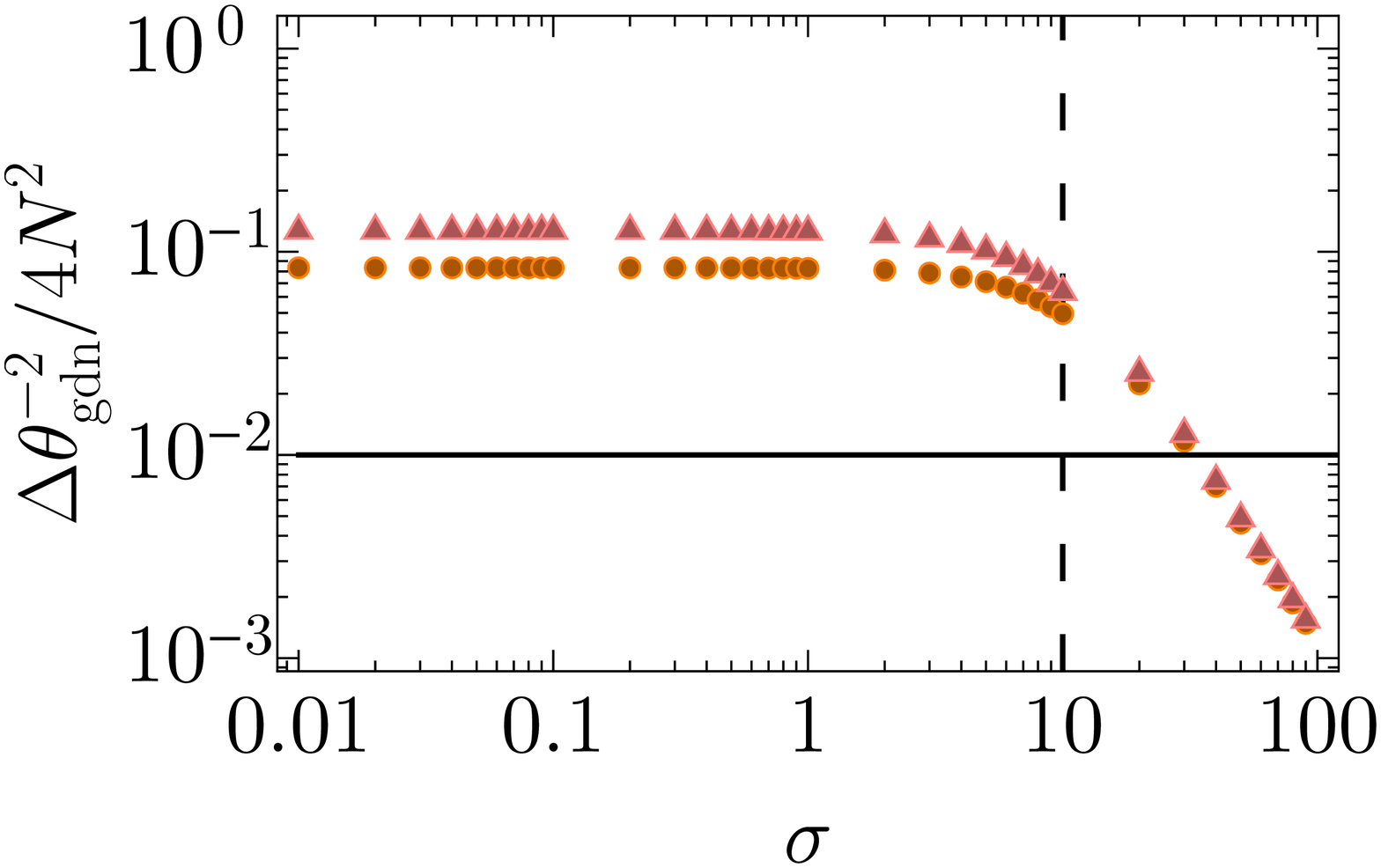}}
\put(165,210){(b)}
% c
\put(-160,0)
{\includegraphics[width=0.3696\linewidth]{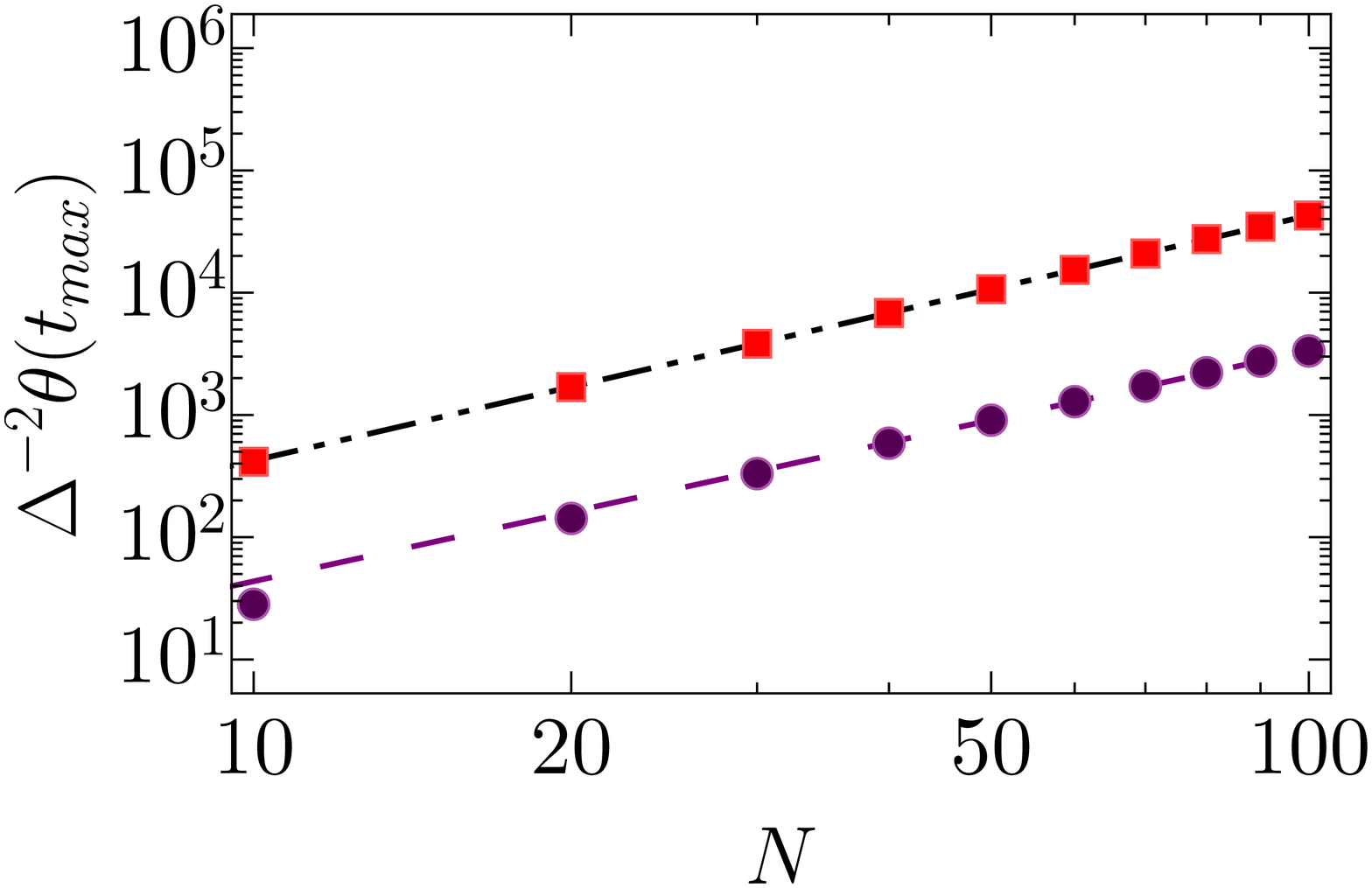}}
\put(-125,90){(c)}
% d
\put(25,0)
{\includegraphics[width=0.3696\linewidth]{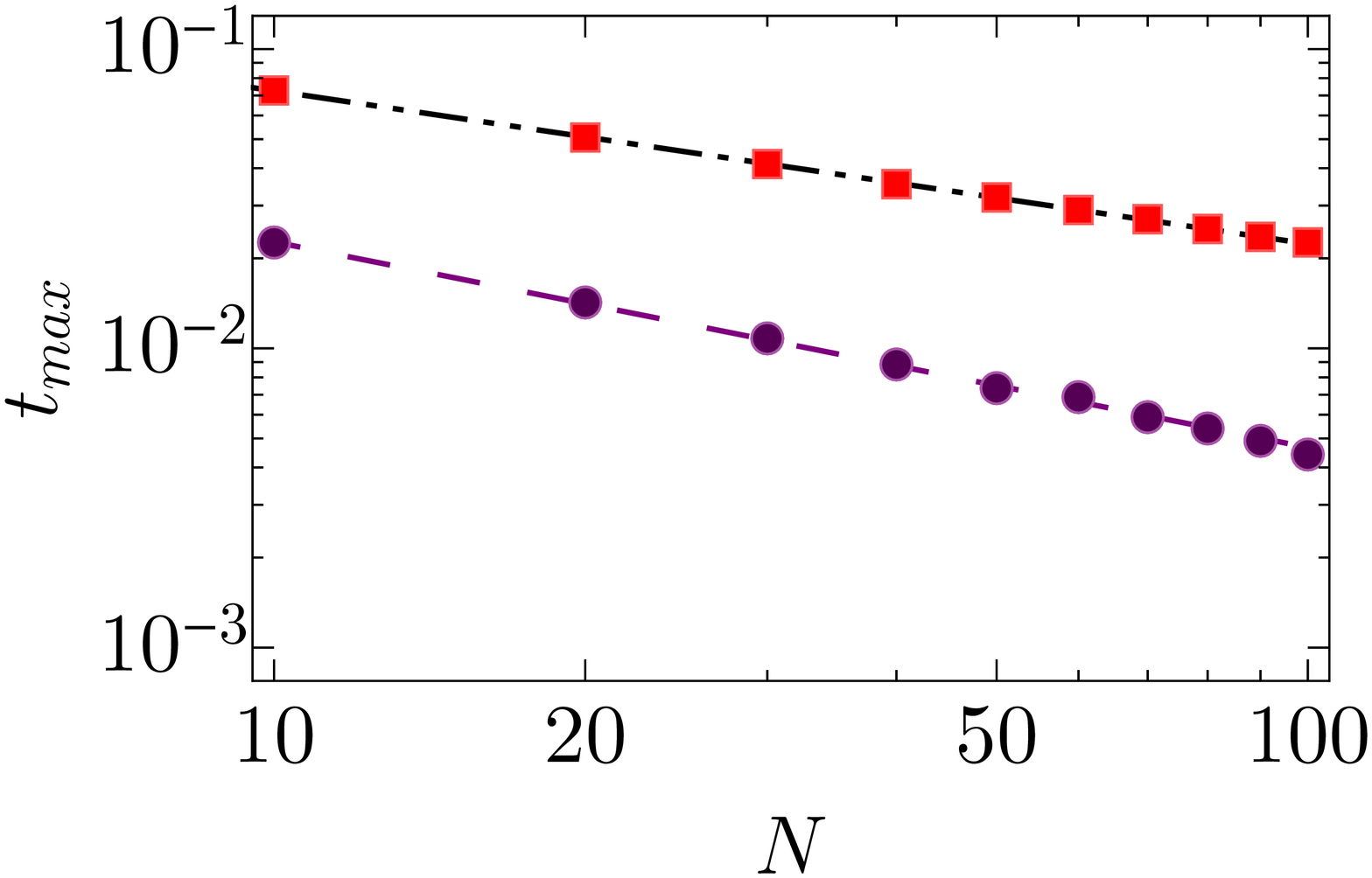}}
\put(165,90){(d)}
\end{picture}
\caption{
(a) $\Delta^{-2} \theta$ from the error-propagation formula (\ref{eq:QyzQYZQFI}) with inverted by the single rotation evolution (\ref{eq:evol_invert}) for $\hat{\mathcal{P}}=\hat{Q}_{yz}$ and $\hat{\Lambda}_{\bf n}^{(III)}=\hat{Q}_{yz}$ is shown by the solid orange line and compared to the ideal interaction-based readout protocol marked by the pink dashed line. 
(b) $\Delta^{-2} \theta^{(I)}_{\rm gdn}(\hat{Q}_{yz})$ versus the detection noise $\sigma$ at $\bar{t}_1=0.01$ for the interaction-based readout protocol (pink triangles) and inversion performed with extra rotation (orange circles). The black solid line marks the position of the SQL and the black dashed vertical line stands for $\sigma=\sqrt{N}$.
(c) Scaling of the first maximum in $\Delta^{-2} \theta^{(I)}(\hat{Q}_{yz})$ versus the total atom number $N$ for the interaction-based readout protocol (red squares) and inverted by the single rotation evolution (purple circles). The fitted exponential function for the ideal interaction-based readout protocol is $f(N)=4.18 N^2$ (red squares and black double dot dashed line) and for the evolution inverted by the rotation is $f(N) = 0.57 N^{1.88}$ (purple circles with the dashed purple line). 
(d) Scaling of the time corresponding to the first maximum of $\Delta^{-2} \theta$ with the total atom number $N$. Fitting with an exponential function gives $f(N)=0.24 N^{-0.51}$ (red squares and the black double dot dashed line) for the interaction-based readout protocol, and $f(N) = 0.11 N^{-0.69}$ (purple circles and the purple dashed line) for inverted by the rotation evolution.}
\label{fig:fig9}
\end{figure}

Therefore, we consider the time evolution of the initial state in the following way:
\begin{equation}\label{eq:evol_invert}
|\psi(\theta, \alpha)\rangle_{\rm invt} = e^{-i \bar{t}_1 \hat{J}^2} e^{-i\alpha \hat{J}_{z, S}} e^{-i\theta \hat{Q}_{yz}}e^{-i \bar{t}_1 \hat{J}^2}|0,N,0\rangle .
\end{equation}
As long as the state is squeezed, or a bit oversqueezed, the angle $\alpha$ can be treated as \cite{FerriniPhd2011}
\begin{equation}\label{eq:alpha}
\tan(2 \alpha)=\frac{\Gamma_{12}}{\Gamma_{11} - \Gamma_{22}},
\end{equation}
where $\Gamma_{ij}$ are covariance matrix elements calculated in Section \ref{sec:identification}. We show numerically, that such a rotation allows us to inverse the evolution at the very early stage, and thus protects the signal against the detection noise without significant lose of information comparing to the ideal situation shown in Fig.~\ref{fig:fig5}. The variation of the corresponding $\Delta^{-2} \theta$ versus the total atom number is $\sim N^2$, while the change of the time scale is typically $t\sim N^{-1/2}$ for the ideal protocols and $t\sim N^{-2/3}$ for the evolution (\ref{eq:evol_invert}) inverted by the rotation, see Fig.~\ref{fig:fig9} for more details.

The interesting question arises if the same trick can be applied in a bimodal system that is more often used for the squeezing generation according to the one-axis model~\cite{PhysRevA.47.5138}. The answer is positive. One can apply the rotation to invert the evolution of a quantum state in the two mode systems as we have checked numerically for the angle $\alpha$ given by the formula $\tan(2\alpha) =(\gamma_{12})/(\gamma_{22}-\gamma_{11})$, where $\gamma_{11}=\Delta^2\hat{S}_x$, $\gamma_{22}=\Delta^2\hat{S}_y$, $\gamma_{12}=-\langle \{\hat{S}_x, \hat{S}_y\} \rangle $, and $\hat{S}_{x,y,z}$ are spin operators defined for the bimodal system. The expressions for $\gamma_{11}$, $\gamma_{22}$ and $\gamma_{12}$ can be calculated analytically~\cite{FerriniPhd2011}. In fact, the numerical results for $\Delta^{-2}\theta$ from the error-propagation formula with $\hat{\Lambda}_n=\hat{S}_y$ and $\hat{\mathcal{P}}=\hat{S}_y$ show that it behaves like the inverse of the squeezing parameter \cite{Wineland1992}. The gain is the resistance of such the squeezing parameter inverse against the detection noise.

%%%%%%%%%%%%%%%%%%%%%%%%%%%%%%%%%%%%%%%%%%%%%%%%%%%%%%%%%%%%%%%%%%%%%%%%%%%%%%%%%%%%%%%%%%%%%%%%%
\section{Summary}

We implemented the quantum interferometry concept in spinor Bose-Einstein condensates employing the time evolved polar state. We focused on the quantum Fisher information in order to identify the best configurations. We solved analytically the dynamics of the polar state in the total spin eigenbasis, paying special attention to quantities that are important to calculate the QFI value. We found out three optimal generators of the  interferometeric rotation that lead to Heisenberg scaling of the QFI, among which two, namely $\hat{\Lambda}^{(II)}_{\mathbf{n}}=\hat{D}_{xy}$ and $\hat{\Lambda}^{(III)}_{\mathbf{n}}=\hat{Y}$, were already successfully implemented experimentally in \cite{Lucke2011} and \cite{OberthalerSU11}, respectively. However, we found out that there is even a better choice for the generator of the interferometric rotation, which is $\hat{\Lambda}^{(I)}_{\mathbf{n}}=\hat{Q}_{yz}$, because it gives much higher value of the QFI in early times of the evolution. Experimental realization of this interferometer with nowadays techniques is possible, although it will require a three-step procedure as explained in Fig.~\ref{fig:fig3}. 

Next, we considered optimal observables that would allow one to exploit the potential of particular interferometers based on the error-propagation formula. Indeed, we established that relatively easy to measure observables $\hat{J}_z^2, \hat{Y}, \hat{Q}_{yz}$ are sufficient to reach the QFI value. However, it turns out that only some of them are resistant against the detection noise even when the interaction-based readout protocol was employed. 

Finally, we showed how to implement the interaction-based readout protocol, which requires inversion of the evolution, by a single rotation of the state. That idea was not reported yet in the literature, and can be applied in the case of a bimodal system as well. As an example, we considered the most prominent configuration with $\hat{\Lambda}^{(I)}_{\mathbf{n}}=\hat{Q}_{yz}$. We showed, that variation of the corresponding precision in the $\theta$ estimation from error-propagation formula, assuming that $ \hat{\mathcal{P}} =\hat{Q}_{yz}$ is measured, has the Heisenberg scaling with the total atom number, it is $N^2$, while the time scale goes like $N^{-2/3}$. The advantage of our purpose is a resistance of the corresponding precision in the $\theta$ estimation against the detection noise.

\section*{Acknowledgments}

We acknowledge discussions with K. Paw\l{}owski and S. Mirkhalaf. This work was supported by the Polish National Science Center Grants DEC-2015/18/E/ST2/00760 and 2017/25/z/st2/03039 under the QuantERA program.

%%%%%%%%%%%%%%%%%%%%%%%%%%%%%%%%%%%%%%%%%%%%%%%%%%%%%%%%%%%%%%%%%%%%%%%%%%%%%%%%%%%%%%%%%%%%%%%%%
\appendix

\section{SMA and time scales}\label{app:SMA}

In the single mode approximation wave functions of atoms in all three Zeeman components are assumed to be identical. It can be justified when the system size is much smaller than the spin healing length $\xi_{\rm sp}=\hbar/\sqrt{2 m c_2 \rho}$, where the density is $\rho=N/V$. It is in the low density regime, up to a few thousand of atoms, when creation of spin domains, vortices, solitons etc. are energetically costly. The vector field can be then replaced by $\hat{\Psi}^T(\mathbf{r}) = \phi(\mathbf{r}) (\hat{a}_{1}, \hat{a}_{0}, \hat{a}_{-1} )^T$ with the SMA wavefunction $\phi(\mathbf{r})$ defining the spatial mode of the spinor BEC. The SMA wavefunction is a solution of the non-linear Schr\"{o}dinger equation \cite{KAWAGUCHI2012253},
\begin{equation}\label{eq:GPE}
\mu \phi(\mathbf{r}) =
\left( -\frac{\hbar^2 \nabla^2}{2m} + \frac{1}{2}m\sum_{\sigma=x,y,z} \omega_\sigma^2 \sigma^2 + c_0 (N-1) |\phi(\mathbf{r})|^2 \right)\phi(\mathbf{r}),
\end{equation}
with constraint $\int d^3 {\bf r} |\phi(\mathbf{r})|^2 = 1$, because $\int d^3{\bf r} \langle \hat{\Psi}^\dagger(\mathbf{r}) \hat{\Psi}(\mathbf{r}) \rangle = N$ and $\sum_{m_F} \langle \hat{a}^\dagger_{m_F} \hat{a}_{m_F} \rangle =N$. Then one obtains the Hamiltonian (\ref{eq:ham}). The energy and time scales are then associated to the SMA wave function $\phi({\bf r})$ through $2c'_2 = c_2 \int d^3r |\phi(\mathbf{r})|^4$.

The wave function $\phi(\mathbf{r})$ can be estimated e.g. in the Thomas-Fermi (TF) approximation, and for $\omega_\sigma=\omega$ one has
\begin{equation}
    c'_2 = \frac{15}{28\pi} \frac{c_2}{r_{tf}^{3}},
\end{equation}
where the TF radius is $r_{tf}^5=\frac{15}{4\pi}\frac{c_0 (N-1)}{m \omega^2}$ and the chemical potential is $\mu =\frac{1}{2}m\omega^2 \left( \frac{15}{4\pi} \frac{c_0 (N-1)}{m \omega^2} \right)^{2/5}$. In what follows, the corresponding time scale 
\begin{equation}
    t_{unit}=\frac{\hbar}{c'_2}=\frac{28 \pi}{15} \frac{\hbar r_{tf}^{3}}{c_2}
\end{equation}
depends on the frequency of the trapping potential and the total atom number. When $\omega=2\pi \times 300$s${}^{-1}$, $N=100$, one obtains $t_{unit} \approx 7$s for sodium atoms when $c_0=4 \pi \hbar^2(2a_2+a_0)/(3m)$, $c_2=4 \pi \hbar^2(a_2-a_0)/(3m)$ and scattering length $a_0=50 a_B$, $a_2=55 a_B$ with the Bohr radius $a_B$~\cite{RevModPhys.85.1191}.

On the other hand, the SMA wave function is a solution of the GPE (\ref{eq:GPE}) corresponding to the lowest energy state, and therefore can be calculated numerically. Both $c'_2$ and the time unit can be calculated numerically as well by integrating that ground state wave function. Then, for the same parameters as considered in the paper we obtained the exact value for the time unit, which is $\hbar /c'_2 \approx 21$s as it was found by us using the imaginary time evolution method.% for the number of grid points $316, 2^3, 48^8$ and time step $dt=10^{-4},10^{-5}$ms and L_x=5-8 r_tf. 
The discrepancy between the TF and exact numerical results is due to the small number of atoms considered. In the small atoms number limit the kinetic energy part is of the order of the interaction term and cannot be neglected. Therefore, the TF approximation is not valid. We checked our statement by calculating numerically the ground state of (\ref{eq:GPE}) with the kinetic energy deleted, and obtained that the numerical result for the time unit is in agreement with the TF approximation. However, the TF approximation provides a fairly good estimate in thermodynamic limit as it was verified by us within exact numerical calculations.

%The time unit is $t_{unit} \approx 12$s for the parameters considered in the paper. The  dimensionless $\bar{t}_1=1$ corresponds to $12$s.

\section{Time evolution in the spin basis}\label{app:evolutionSPIN}

The Hamiltonian (\ref{eq:ham}) is diagonal in the spin basis $|N,J,M\rangle$ whose standard representation is
\begin{equation}\label{eq:spin_state}
|N,J,M\rangle = \frac{1}{\sqrt{\mathcal{Z}(N,J,M)}}(\hat{J}_-)^{P} (\hat{A}^{\dagger})^{Q}(\hat{a}_{+1})^J|0\rangle,
\end{equation}
with $P=J-M$, $2Q=N-J$, $\hat{J}_{-} = \sqrt{2}(\hat{a}_{-1}^{\dagger}\hat{a}_0 + \hat{a}_{0}^{\dagger}\hat{a}_{+1})$, $\hat{A}^{\dagger} = (\hat{a}_{0}^{\dagger})^2 - 2\hat{a}_{+1}^{\dagger}\hat{a}_{-1}^{\dagger}$ and $|0\rangle$ is the vacuum state. The state $|N,J,M\rangle$ is parametrized by three quantum numbers: the total atom number $N$, the total spin length $J$ which take integer values with the same parity as $N$ to makes $2Q$ even, and magnetization $M \in \{ -N, -N+1, \ldots, N\}$. The normalization factor $\mathcal{Z}(N,S,M)$ reads
\begin{equation}
\mathcal{Z}(N,J,M) = \frac{J!(N-J)!!(N+J+1)!!(J-M)!(2J)!}{(2J+1)!!(J+M)!}.
\end{equation}

The spin states $|N,J,M\rangle$ are simultaneous eigenstates of the $\hat{J}^2$ and $\hat{J}_z$ operators:
\begin{align}
	\hat{J}^2 |N,J,M\rangle & = J(J+1)|N,J,M\rangle, \\
	\hat{J}_z |N,J,M\rangle & = M|N,J,M\rangle.
\end{align}

The spin state (\ref{eq:spin_state}) can be decomposed in the Fock state basis $|N_{1}, N_0, N_{-1}\rangle \equiv |l,N+M-2l,l-M\rangle$ which we parametrized by the single parameter $l$ because of the fixed total atom number $N=N_1+N_0+N_{-1}$ and magnetization $M=N_1-N_{-1}$. By using the general definition  (\ref{eq:spin_state}) we obtained

\begin{equation}\label{eq:fock_expansion}
|N,J,M\rangle = \frac{1}{\sqrt{\mathcal{Z}(N,J,M)}}\sum\limits_{l=l_{\rm min}}^{l_{\rm max}} D_{l}(N,J,M)|l,N+M-2l,l-M\rangle,
\end{equation}
where
\begin{align}
&D_{l}(N,J,M) = \nonumber \\
&W_l(N,J,M)
	\sum\limits_{n=0}^{\lfloor\frac{J-M}{2} \rfloor}\frac{2^{-n}}{n!}  \sum\limits_{k=0}^{\frac{N-J}{2}} \frac{(-2)^k (k+J)!(N-J-2k)!}{(k-l-n+J)!(l-k-n-M)!(N+M-J+n-k-l)!}\binom{\frac{N-J}{2}}{k},
\end{align}
in which
\begin{equation}
W_l(N,J,M)=(J-M)!2^{\frac{J-M}{2}}\sqrt{\frac{(l-M)!(N+M-2l)!}{l!}},
\end{equation}

and $l_{\rm min} = \min(0,M)$ and $l_{\rm max} = \lfloor(N+M)/2\rfloor$. 

The initial polar state $|0,N,0\rangle$ can be decomposed in the spin basis using Eq.~\eqref{eq:fock_expansion}, and it takes the form
\begin{equation}
	|0,N,0\rangle = \sum\limits_{J=0}^N{}^{'} D_{0}(N,J,0)|N,J,M=0\rangle,
\end{equation}
in which the coefficient $D_{0}(N,J,0)$ has a closed form expression
\begin{equation}
	D_{0}(N,J,0) = \sqrt{\binom{\frac{N+J}{2}}{J} \binom{N+J}{J}^{-1} \frac{(2J+1)}{(N+J+1)}2^J} 
	\simeq e^{-\frac{J(J-1)}{4 N}} \sqrt{\frac{2J+1}{N+J}}.
\end{equation}

Bringing everything together, the time evolution of the polar state, $|\psi(t)\rangle = \exp(-i \bar{t} \hat{J}^2)|0,N,0\rangle$, is 
\begin{equation}\label{eq:evolution_angular_mom}
	|\psi(t)\rangle = \sum\limits_{J=0}^N{}^{'} D_0(N,J,0) e^{-i\bar{t}J(J+1)}|N,J,M=0\rangle,
\end{equation}
and demonstrates that the dynamics is periodic with period $\Delta \bar{t} = \pi$.

Once we treated the time evolution of the polar state analytically, the evolution of particular quantities of interest can also be calculated analytically using relations shown in~\ref{app:useful}, \ref{app:aanihiation} and \ref{app:anumber}. The final results are quite complex, however they can be expressed by a single sum over the total spin length $J$ as shown in Section~\ref{sec:intro}.

\section{Some useful relations needed for derivation of actions of annihilation and particle number operators onto the spin states}\label{app:useful}

\begin{align}
\comm{\hat{a}_{m_F}}{\hat{a}_{m_F}^{\dagger}{}^{N}}&=N\hat{a}_{m_F}^{\dagger}{}^{N-1}, \\
\comm{\hat{a}_{m_F}^{\dagger}}{\hat{a}_{m_F}^{N}}&=-N\hat{a}_{m_F}^{N-1},\\
\comm{\hat{a}_{0}}{\hat{J}_{-}^{P}}&=\sqrt{2}P\hat{J}_{-}^{P-1}\hat{a}_{+1},\\
\comm{\hat{a}_{+1}}{\hat{J}_{-}^{P}}&=0,\\
\comm{\hat{a}_{-1}}{(\hat{A}^{\dagger})^{Q}}&=-2Q(\hat{A}^{\dagger})^{Q-1}\hat{a}^{\dagger}{}_{+1},\\
\comm{\hat{a}_{0}}{(\hat{A}^{\dagger})^{Q}}&=2Q(\hat{A}^{\dagger})^{Q-1}\hat{a}^{\dagger}{}_{0},\\
\comm{\hat{a}_{+1}}{(\hat{A}^{\dagger})^{Q}}&=-2Q(\hat{A}^{\dagger})^{Q-1}\hat{a}^{\dagger}{}_{-1},\\
\comm{\hat{A}^{\dagger}}{\hat{J}_{-}}&=0,\\
\comm{\hat{a}_{-1}}{\hat{J}_{-}^{P}}&=\sqrt{2}P\hat{J}_{-}^{P-1}\hat{a}_{0}+P(P-1)\hat{J}_{-}^{P-2}\hat{a}_{+1}
\end{align}
and also
\begin{align}
	\hat{a}^{\dagger}_{0}(\hat{a}^{\dagger}_{+1})^{J}\ket{0}&=\frac{1}{\sqrt{2}(J+1)}\hat{J}_{-}(\hat{a}^{\dagger}_{+1})^{J+1}\ket{0},\\
	\hat{a}^{\dagger}_{-1}(\hat{a}^{\dagger}_{+1})^{J}\ket{0}&=\frac{1}{2(J+1)(1+2J)}(\hat{J}_{-})^{2}(\hat{a}^{\dagger}_{+1})^{J+1}\ket{0}\nonumber \\
	&-\frac{J}{1+2J}\hat{A}^{\dagger}(\hat{a}^{\dagger}_{+1})^{J-1}\ket{0}.
\end{align}

\section{Action of annihilation operators on the spin state}\label{app:aanihiation}
\begin{align}
	&\hat{a}_{+1}|N,J,M\rangle  = -\sqrt{B_{+}(N,J,M)}|N-1,J+1,M-1\rangle
	+ \sqrt{B_{-}(N,J,M)}|N-1,J-1,M-1\rangle,
\\
	&\hat{a}_{0}|N,J,M\rangle = \sqrt{A_{+}(N,J,M)}|N-1,J+1,M\rangle
	+ \sqrt{A_{-}(N,J,M)}|N-1,J-1,M\rangle,
\\
&\hat{a}_{-1}|N,J,M\rangle = -\sqrt{B_{+}(N,J,-M)}|N-1,J+1,M+1\rangle
	+  \sqrt{B_{-}(N,J,-M)}|N-1,J-1,M+1\rangle, 
\end{align}
where
\begin{equation}
	A_{+}(N,J,M) = \frac{(N-J)(J-M+1)(J+M+1)}{(1+2J)(2J+3)} ,
\end{equation}
\begin{equation}
	A_{-}(N,J,M) = \frac{(N+J+1)}{(2J+1)(2J-1)} (J-M)(J+M),
\end{equation}
\begin{equation}
	B_{+}(N,J,M) = \frac{(N-J)(J-M+1)(J-M+2)}{2(1+2J)(2J+3)} ,
\end{equation}
\begin{equation}
	B_{-}(N,J,M) = \frac{(N+J+1)(J+M)(J+M-1)}{2(2J+1)(2J-1)} .
\end{equation}

\section{Action of particle number operators on the spin state}\label{app:anumber}

\begin{align}
	&\hat{N}_{+1}|N,J,M\rangle = (B_{+}(N,J,M) + B_{-}(N,J,M))|N,J,M\rangle \nonumber\\
	&- \sqrt{B_{+}(N,J,M)B_{-}(N,J+2,M)}|N,J+2,M\rangle
	- \sqrt{B_{-}(N,J,M)B_{+}(N,J-2,M)}|N,J-2,M\rangle,
\\
	&\hat{N}_{0}|N,J,M\rangle = (A_{+}(N,J,M) + A_{-}(N,J,M))|N,J,M\rangle \nonumber \\ &+\sqrt{A_{+}(N,J,M)A_{-}(N,J+2,M)}|N,J+2,M\rangle 
	+ \sqrt{A_{-}(N,J,M)A_{+}(N,J-2,M)}|N,J-2,M\rangle,
\\
	&\hat{N}_{-1}|N,J,M\rangle = (B_{+}(N,J,-M) + B_{-}(N,J,-M))|N,J,M\rangle \nonumber \\ &-\sqrt{B_{+}(N,J,-M)B_{-}(N,J+2,-M)}|N,J+2,M\rangle \nonumber\\
	& - \sqrt{B_{-}(N,J,-M)B_{+}(N,J-2,-M)}|N,J-2,M\rangle. 
\end{align}

%%%%%%%%%%%%%%%%%%%%%%%%%%%%%%%%

\section{Undepleted pump approximation}\label{app:undepleted}

In the case when the evolution is starting from the macroscopically populated $m_F=0$ mode, it is tempting to examine the undepleted pump approximation where the mode $m_F=0$ acts as a source and injects atoms to the side modes. When the number of atoms in the $m_F=0$ component is close to $N$, i.e. initially and for short times, one can replace the annihilation operator $\hat{a}_0 \to \sqrt{N} e^{-i\chi_p/2}$. It means that, the mode $m_F=0$ is decoupled from $m_F=\pm 1$ and has a constant occupation $\langle \hat{N}_0 \rangle = N$. The Hamiltonian takes then the form
\begin{equation}
\ham = 2c'_2 \left[\left(N-\frac{1}{2}\right)(\hat{N}_{+1} + \hat{N}_{-1}) + N e^{-i\chi_{p}}\hat{a}^{\dagger}_{+1}\hat{a}^{\dagger}_{-1} + N e^{i\chi_{p}}\hat{a}_{+1}\hat{a}_{-1} \right].
\end{equation}
The Heisenberg equation of motion reads
\begin{equation}
\frac{d}{dt} \hat{a}_{\pm 1}(t) = \frac{i}{\hbar}[\ham, \hat{a}_{ \pm 1}] = -\frac{2i c'_2}{\hbar}\left[ \left(N - \frac{1}{2} \right) \hat{a}_{\pm 1}(t) + e^{-i\chi_{p}}N \hat{a}^{\dagger}_{\mp 1}(t)\right].
\end{equation}
The evolution of annihilation operators can be found exactly~\cite{PhysRevA.77.023616,PhysRevLett.115.163002}:
\begin{align}
	\hat{a}_{\pm 1}(t) & = \mathcal{A}(t) \hat{a}_{\pm 1}(0) + \mathcal{B}(t) \hat{a}_{\mp 1}(0), \\[3mm]
	\mathcal{A}(t) & = \cosh\left(\frac{c'_2 t}{\hbar}\sqrt{4 N - 1} \right) - i \frac{(2 N - 1)}{\sqrt{4 N- 1}}\sinh\left(\frac{c'_2 t}{\hbar}\sqrt{4N - 1}  \right) \\[3mm]
	\mathcal{B}(t) & = -\frac{2i e^{-i\chi_{p}}N}{\sqrt{4 N-1}}\sinh\left(\frac{c'_2 t}{\hbar}\sqrt{4 N - 1}  \right),
\end{align}
leading to the rapid grow of the total mean occupation of side modes
\begin{equation}
	\langle \hat{N}_s \rangle = \frac{8 N^2}{4 N-1}\sinh^2\left(\frac{c'_2 t}{\hbar}\sqrt{4N-1} \right),\label{eq:AFNs}
\end{equation}
and the variance fully determined by the mean occupation, i.e. 
\begin{equation}
	\Delta^2 \hat{N}_s  = \langle \hat{N}_s \rangle(\langle \hat{N}_s \rangle + 2).\label{eq:AFNs2}
\end{equation}

\begin{figure}[]
    \begin{center}
	\includegraphics[width=0.6\linewidth]{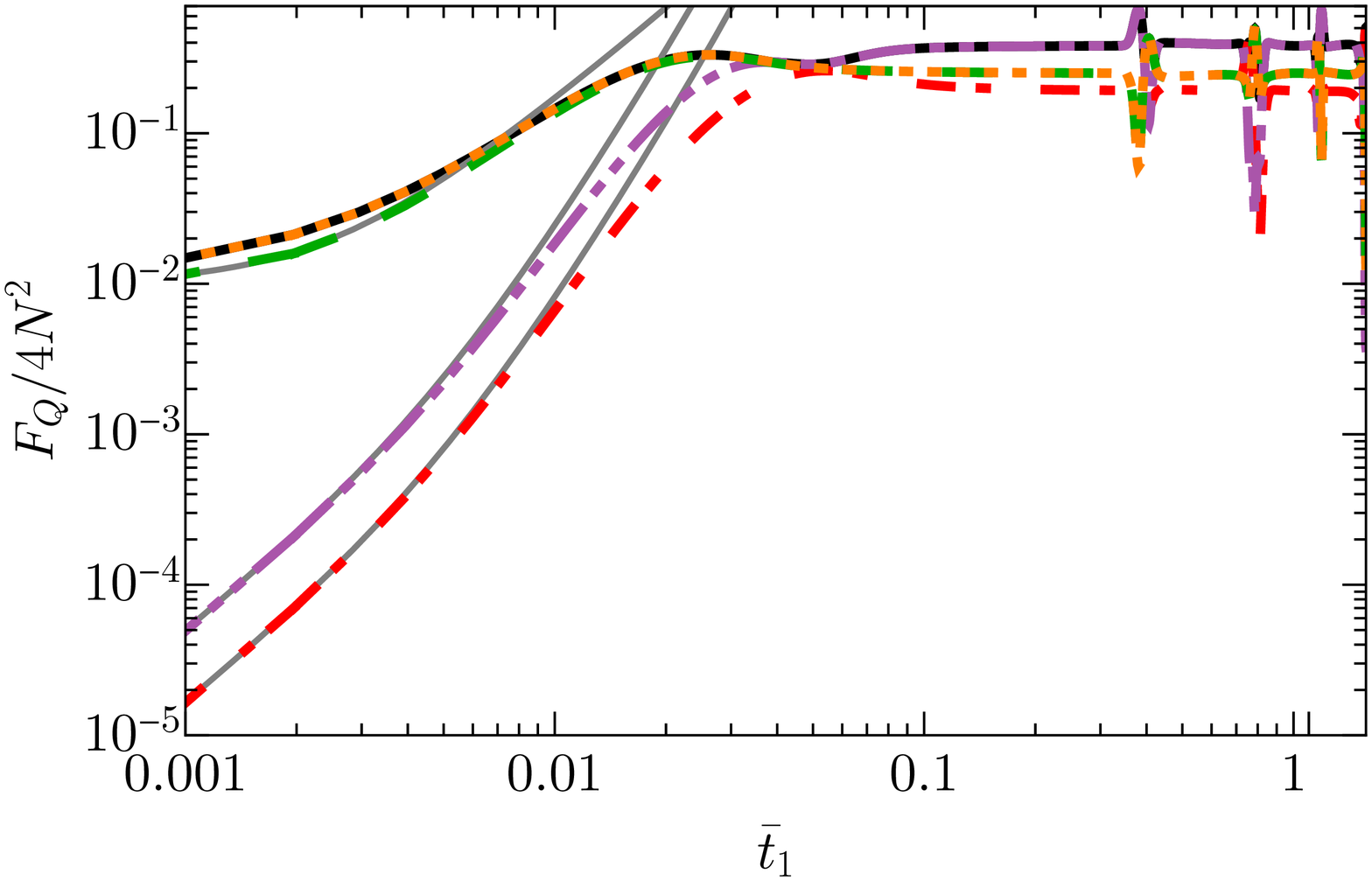}
	\end{center}
	\caption{
	The same as in Fig.\ref{fig:fig2} but with the logarithmic scale on the vertical axis. 	Variations of the QFI in time for different interferometers $\hat{\Lambda}^{(I)}_{\mathbf{n}}= \hat{Q}_{yz}$ (dashed green line), $\hat{\Lambda}^{(II)}_{\mathbf{n}}=\hat{D}_{xy}$ (dot dashed red line) and $\hat{\Lambda}^{(III)}_{\mathbf{n}}=\hat{Y}$ (double dot dashed violet line). In addition, $\hat{\Lambda}^{(I)}_{\mathbf{n}}=\hat{J}_x(\gamma)$, with $\gamma$ chosen such that it maximizes the QFI value, is also shown by the orange dotted line for comparison. The  QFI maximized over all interferometers is shown by the black solid line. 
	The corresponding variations of QFI from the undepleted pump approximation are marked by the solid gray lines. The corresponding QFI from the approximation always overestimates its exact value.}	\label{fig:figAF}
\end{figure}

The variances of relevant operators that determine the value of quantum Fisher information can be expressed in terms of (\ref{eq:AFNs}) and (\ref{eq:AFNs2}), and they are
\begin{align}
	F_Q^{(I)}=4\Delta^2 \hat{Q}_{yz} & = 4\langle\hat{N}_0 (2\hat{N}_s + 1) \rangle = 4(2 \langle \hat{N}_s\rangle + 1) N, \\
	F_Q^{(II)}=4\Delta^2 \hat{D}_{xy} & = 2\langle \hat{N}_s (\hat{N}_s + 2) \rangle = 4\langle \hat{N}_s \rangle(\langle \hat{N}_s \rangle + 2), \\
	F_Q^{(III)}=4\Delta^2 \hat{Y} & = 12 \langle (\Delta \hat{N}_{s})^2\rangle = 12 \langle\hat{N}_{s} \rangle(\langle\hat{N}_{s} \rangle + 2).
\end{align}
The undepleted pump approximation overestimates the true value of the QFI as demonstrated in Fig.\ref{fig:figAF}, and the difference is apparent very quickly. The approximation is fairly good up to $\bar{t} \lesssim 0.1/\sqrt{N}$.

\section{Mean-field phase portraits in the symmetric subspace}\label{app:meanfield}

The parametrization of the symmetric and anti-symmetric spin operators in both symmetric and anti-symmetric subspaces is
\begin{align}
    &\hat{J}^{cl}_{x,l}=N\sin{\tilde{\theta}}\cos{\tilde{\phi}} \\
    &\hat{J}^{cl}_{y,l}=N\sin{\tilde{\theta}}\sin{\tilde{\phi}} \\
    &\hat{J}^{cl}_{z,l}=N\cos{\tilde{\theta}},
\end{align}
where $\tilde{\phi} \in [0,2\pi]$ is the azimuthal angle and $\tilde{\theta} \in [0,\pi]$ is the polar angle of the Bloch sphere. The total energy on the mean field level in the symmetric subspace is
\begin{align}\label{eq:energy}
 E_S= J_{x, S}^{cl}{}^2 + J_{y, A}^{cl}{}^2 = N^2 \sin^2(\tilde{\theta}),
\end{align}
and so equations of motion are given by
\begin{align}\label{eq:motion}
    \dot{\tilde{\phi}}=\frac{2}{\hbar}\frac{\partial E}{\partial \tilde{\theta}}=N^2 \sin{2 \tilde{\theta}}, \\
    \dot{\tilde{\theta}}=-\frac{2}{\hbar}\frac{\partial E}{\partial \tilde{\phi}}=0.
\end{align} 

\section*{References}

%\bibliography{bibliography}

\providecommand{\newblock}{}

\end{document}